\documentclass[amsmath,amssymb,twocolumn]{revtex4}
\usepackage{graphicx}
\usepackage{bm}           

\usepackage[usenames]{color}
\usepackage{cancel}
\usepackage{hyperref}
\usepackage[normalem]{ulem}

\usepackage{dcolumn}
\usepackage{bm}
\usepackage{array}
\usepackage{color}

\begin{document}

\title{Feedback arcs and node hierarchy in directed networks}

\author{
  Jin-Hua Zhao$^{1}$\footnote{Current address: Department of Applied Science and Technology, Politecnico di Torino, Italy} and
  Hai-Jun Zhou$^{1,2}$\footnote{Corresponding author.
    Email: {\tt zhouhj@itp.ac.cn}}
}

\affiliation{
  $^1$Key Laboratory of Theoretical Physics, Institute of Theoretical Physics, Chinese Academy of Sciences, Zhong-Guan-Cun East Road 55, Beijing 100190, China \\
  $^2$School of Physical Sciences, University of Chinese Academy of Sciences, Beijing 100049, China
}

\date{16 December, 2016}

\begin{abstract}
  Directed networks such as gene regulation networks and neural networks are connected by arcs (directed links). The nodes in a directed network are often strongly interwound by a huge number of directed cycles, which lead to complex information-processing dynamics in the network and make it highly challenging to infer the intrinsic direction of information flow. In this theoretical paper, based on the  principle of minimum-feedback, we explore the node hierarchy of directed networks and distinguish feedforward and feedback arcs. Nearly optimal node hierarchy solutions, which minimize the number of feedback arcs from lower-level nodes to higher-level nodes, are constructed by belief-propagation and simulated-annealing methods. For real-world networks, we quantify the extent of feedback scarcity by comparison with the ensemble of direction-randomized networks and identify the most important feedback arcs. Our methods are also useful for visualizing directed networks.
\end{abstract}

\maketitle

\section{Introduction}

Directed networks are formed by nodes and arcs (i.e., directed links) pointing from one node to another. They are ubiquitous in biological and technological systems; for instance, neurons in the brain rely on directed synaptic connections to form an information-processing network \cite{White-etal-1986}, and cell regulatory networks contain directed interactions between genes, proteins and other small molecules \cite{Li-etal-2004,Oda-etal-2005,Alon-2007}. Structural properties of directed networks at different scales have been studied in the literature for many years, especially on network small  motifs \cite{Alon-2007,Milo-etal-2002}, mesoscopic communities \cite{Leicht-Newman-2008,Fortunato-2010}, strongly connected components \cite{Tarjan-1972,Dorogovtsev-etal-2001}, and network hierarchical structure \cite{Jeong-etal-2000,Ravasz-etal-2002,Lan-Mezic-2011,CorominasMurtra-etal-2013,DominguezGarcia-Pgolotti-Munoz-2014}. A directed network can easily be decomposed into a set of strongly connected components (SCCs) and at this coarse-grained level is a directed acyclic feedforward graph of SCCs, with clear-cut hierarchical structure as directed cycle is absent \cite{Dorogovtsev-etal-2001,CorominasMurtra-etal-2013}. Each SCC is itself a maximal subnetwork formed by some nodes and the arcs between them, and any node can reach and be reached by any another node of the same SCC through at least one directed path. Directed cycles are usually abundant in the large SCCs (each of which contains many nodes and arcs), and they cause strong feedback effect and make the information-processing dynamics in the network highly complex \cite{Fiedler-etal-2013,Xu-Lan-2015}.

The hierarchical structures within large SCCs of directed networks have not yet been fully investigated except for a few earlier efforts (e.g., \cite{Lan-Mezic-2011,DominguezGarcia-Pgolotti-Munoz-2014,Xu-Lan-2015}). Due to the cyclic nature of a SCC, it appears at first sight to be quite ambiguous or even meaningless to order its nodes in a particular way and to define an intrinsic flow direction \cite{CorominasMurtra-etal-2013}. However in this paper we show that the arcs that are most vital for feedback interactions can be idenified by collectively considering all the directed cycles of the network. We take an optimization approach based on the so-called principle of minimum feedback \cite{Lan-Mezic-2011}, which defines the minimum feedback arc set problem. An integer hierarchical level is assigned to each node of the input network and the resulting level configuration of all the nodes is called a node hierarchy. The node levels in this hierarchy are optimized by two efficient physics-inspired algorithms, SA and BPD, which minimize the total number of feedback arcs (defined as those pointing from lower-level nodes to higher-level nodes).

\begin{figure}[t]
  \begin{center}
    \includegraphics[angle=0, width=1.0\linewidth]{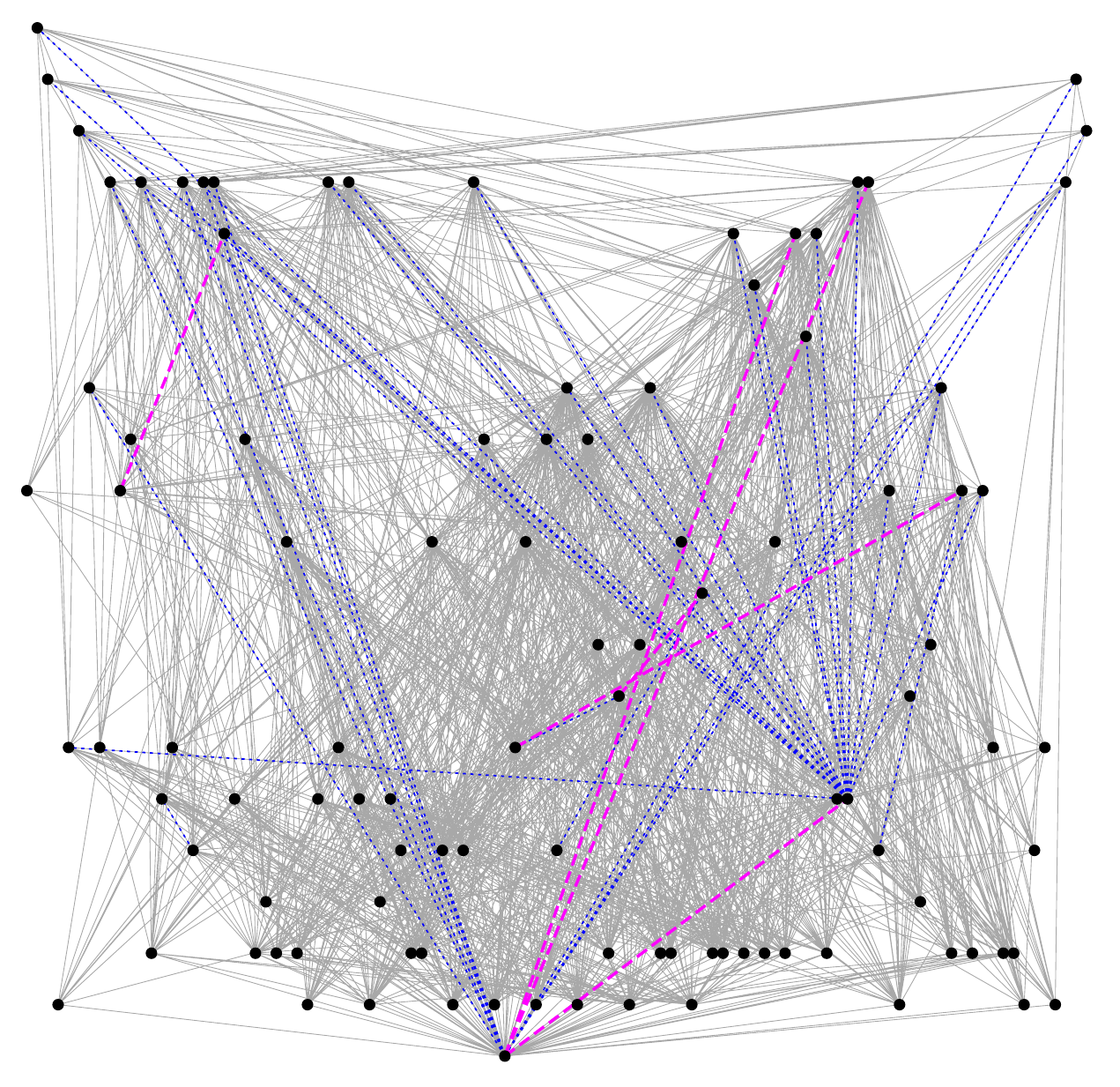}
  \end{center}
  \caption{
    \label{fig:Florida}
    An optimal node hierarchy for the Florida food web \cite{Ulanowicz-etal-1998} and the corresponding feedback arcs. The whole network has $128$ nodes and $2106$ arcs, and its largest strongly connected component contains $103$ nodes and $1579$ arcs  (for clarity only this component is shown). The nodes (black dots) are arranged to $22$ hierarchical levels starting from level $0$ at the bottom. Gray solid lines are feedforward arcs (pointing from higher-level nodes to lower-level nodes), red dashed lines are feedback arcs (pointing from lower-level nodes to higher-level nodes). Each blue dotted line represents a pair of opposite arcs between two nodes.
  }
\end{figure}

Given a real-world directed network, we can construct many near-minimum feedback arc sets by repeatedly running the SA or BPD algorithm. The sizes of these constructed sets are very close to each other and are much smaller than the total number of arcs in the network. We also find that, while most of the arcs of the network never appear in any of these feedback arc sets, a few of them appear in almost all of them. As a concrete example, for the Florida food web \cite{Ulanowicz-etal-1998} formed by $128$ node and $2106$ arcs, only six of the arcs need to be classified as feedbacks (Fig.~\ref{fig:Florida}), which is much lower than the expected number of $601$ feedback arcs in a direction-randomized network. Our algorithms reveal that two arcs of the food-web network are present in all the minimum feedback arc sets. Similar results are obtained for other real-world networks.

By distinguishing feedforward arcs and feedback arcs for a real-world directed network, our work help to reveal the hidden principle direction of flows in the network and the hierarchical organization of the nodes within the strongly connected network components. For biological networks, the identified most important feedback arcs might serve as optimal targets of intervening the system \cite{Liu-Barabasi-2016}. Our algorithms are also useful for network visualization \cite{Eades-Sugiyama-1990}. The source codes of these algorithms will be publicly available to facilitate analyzing and visualizing biological, technological, and social networks.

\section{Node hierarchy and belief propagation}

Given a directed network $G$ of $N$ nodes and $M$ arcs, with arc density $\alpha \equiv M/N$, we introduce a node hierarchy $\underline{h} \equiv (h_1, h_2, \ldots, h_N)$ to partially order the $N$ nodes $i\in \{1, 2, \ldots, N\}$. The level of each node $i$ takes a nonnegative integer value $h_i \in \{0, 1, \ldots, N-1\}$, and level $0$ is the lowest. A node $j$ at positive level $h_j$ must have at least one outgoing arc $(j, k)$ to a node $k$ at one level below (i.e., $h_k=h_j-1$) to justify its level. Under these level constraints, our goal is to construct an optimal node hierarchy which agrees with most of the arcs, i.e., the total number of arcs from higher-level nodes to lower-level nodes reaches the global maximum value.

The node hierarchy problem is essentially equivalent to the feedback arc set problem, a fundamental and famous non-deterministic polynomial hard (NP-hard) problem in computer science \cite{Garey-Johnson-1979} (Appendix~\ref{sec:appMap}). We can treat the node hierarchy problem as a statistical mechanical system. Let us define the energy of an arc $(i, j)$ as $E_{i j}(h_i, h_j) = 0$ for $h_i > h_j$ and $=1$ for $h_i \leq h_j$. The total energy of hierarchy $\underline{h}$ is then the sum of arc energies,
\begin{equation}
  E(\underline{h}) \equiv \sum\limits_{(i, j)\in G} E_{i j}(h_i, h_j) \; .
\end{equation}

We write down the following equilibrium partition function $Z$ to combine the effects of energy and level constraints:
\begin{equation} 
  \label{eq:modelR}
  Z(\beta) = \sum\limits_{h_1 \geq 0} \cdots \sum\limits_{h_N\geq 0}
  \prod\limits_{i=1}^{N} \psi_i \prod\limits_{(j, k)\in G} \psi_{j k} \; .
\end{equation}
Here $\psi_i$ is the Boltzmann factor of node $i$ due to its level constraint: $\psi_i=1$ if $h_i=0$ or $i$ has an outgoing arc $(i, j)$ to a node $j$ at level $h_j=h_i-1$; otherwise $\psi_i=0$. The Boltzmann factor of arc $(j, k)$ is $\psi_{j k}=1$ if its energy is zero ($h_j > h_k$); otherwise $\psi_{j k} = e^{-\beta}$ with the inverse temperature $\beta$ being an adjustable parameter. Notice that each node hierarchy $\underline{h}$ contributes a weight $e^{-\beta E(\underline{h})}$ to $Z$. At sufficiently large values of $\beta$, the node hierarchies with the global minimum energy value (i.e., the optimal node hierarchies) will have overwhelming contributions to the partition function $Z$.

We have solved model (\ref{eq:modelR}) by the replica-symmetric (RS) cavity method developed in the spin glass research field \cite{Mezard-Montanari-2009,Mezard-Parisi-2001,Bayati-etal-2008,Altarelli-Braunstein-DallAsta-Zecchina-2013,Guggiola-Semerjian-2015,Zhou-2016b} (Appendix~\ref{sec:appModelR}). Due to the strong level constraints the mean-field equations of this RS theory are very complicated and are computationally inefficient.

A set $\Lambda$ of arcs is regarded as a feedback arc set (FAS) if it intersects with every directed cycle of the network. Notice that if all the arcs of a FAS are deleted the remaining network contains no directed cycle. A FAS is a minimal one if any of its proper subset is no longer a FAS; and it is a minimum one if its cardinality is the smallest among all the feedback arc sets.  Given a node hierarchy $\underline{h}$ of network $G$, the set formed by all the arcs $(i, j)$ with $h_i \leq h_j$ is a FAS. On the other hand, a unique node hierarchy can be constructed for any FAS by first deleting all the arcs of this set from the network and assigning the lowest level $0$ to all the nodes which have no outgoing arc, followed by iteratively assigning the level $1, 2, \ldots$ to all the remaining nodes which have outgoing arcs only to nodes at lower levels. Indeed there is a one-to-one correspondence between node hierarchies and the so-called \emph{neat} feedback arc sets (Appendix~\ref{sec:appMap}). All the minimal and minimum feedback arc sets (and some special non-minimal ones) are neat, and therefore an optimal node hierarchy is equivalent to a minimum FAS.

This equivalence means that we can obtain a near-optimal node hierarchy by first constructing a near-minimum FAS. For the latter task the level constraints of Eq.~(\ref{eq:modelR}) are not necessary, so we can drop them by setting the Boltzmann factors of all the nodes $i$ to be $\psi_i\equiv 1$. The RS mean-field theory for this relaxed model is much more convenient for numerical treatment (Appendix~\ref{sec:appModelE}). This simplified theory estimates the probability $\rho_{i j}$ of arc $(i, j)$ being a feedback arc to be
\begin{equation}
  \label{eq:rhoarc}
  \rho_{i j} =
  \frac{e^{-\beta}\sum\limits_{h_i =0}^{D-1} \sum\limits_{h_j=h_i}^{D-1}
    q_{i\rightarrow j}^{h_i} q_{j\rightarrow i}^{h_j}}
       {1- (1-e^{-\beta}) \sum\limits_{h_i = 0}^{D-1}
         \sum\limits_{h_j=h_i}^{D-1} q_{i\rightarrow j}^{h_i}
         q_{j\rightarrow i}^{h_j}}
       \; ,
\end{equation}
where the integer $D$ restricts the level of each node $i$ to be $h_i < D$ to compensate for the removed level constraints; the function $q_{j\rightarrow j^\prime}^{h_j}$ denotes the probability that node $j$ will be at level $h_j$ if node $j^\prime$ is absent. The self-consistent belief propagation (BP) equation for this cavity probability is
\begin{eqnarray}
  \label{eq:BP}
  q_{j \rightarrow j^\prime}^{h_j}   \propto &
  \prod\limits_{i\in p(j)\backslash j^\prime} \Bigl[ 1-(1-e^{-\beta})
    \sum\limits_{h_i = 0}^{h_j} q_{i\rightarrow j}^{h_i} \Bigr] \times 
  \nonumber \\
  & 
  \prod\limits_{k\in c(j)\backslash j^\prime} \Bigl[
    1-(1-e^{-\beta})\sum\limits_{h_k = h_j}^{D-1}
    q_{k\rightarrow j}^{h_k} \Bigr]
  \; ,
\end{eqnarray}
where $p(j) \equiv \{i : (i, j)\in G\}$ and $c(j) \equiv \{k : (j, k)\in G\}$; and $p(j)\backslash j^\prime$ is the subset of $p(j)$ with $j^\prime$ being excluded, similarly for $c(j)\backslash j^\prime$.

We can iterate the BP equation (\ref{eq:BP}) on the network $G$ at a fixed large value of $D$ (e.g., $D=200$) and different values of $\beta$ and then estimate the mean fraction $\rho$ of feedback arcs as
\begin{equation}
  \label{eq:rho00}
  \rho= \frac{1}{M} \sum\limits_{(i, j)\in G} \rho_{i j} \; .
\end{equation}
Based on Eqs.~(\ref{eq:rhoarc}) and (\ref{eq:BP}), a belief-propagation--guided decimation (BPD) algorithm is also implemented to construct near-minimum feedback sets (Appendix~\ref{sec:appBPD}). Briefly speaking, at each decimation step a tiny fraction of arcs $(i, j)$ with the largest estimated $\rho_{i j}$ values are deleted from the network $G$; then $G$ is further simplified by deleting all the nodes which have no incoming or outgoing arc; then Eq.~(\ref{eq:BP}) is iterated a small number of times and the value of $\rho_{k l}$ for each remaining arc $(k, l)$ is updated.

\section{Simulated annealing}
\label{sec:sa}

Let us represent an $N$-node permutation as a column vector $\mathcal{P}\equiv (v_1, v_2, \ldots, v_N)^T$ with $v_r \in \{1, 2, \ldots, N\}$ and $v_r \neq v_{r^\prime}$ if $r\neq r^\prime$. Another way of simplifying the level constraints of Eq.~(\ref{eq:modelR}) is to set the level $h_i$ of each node $i$ to be its vertical position in $\mathcal{P}$. A most convenient way of permutating the nodes to reduce the total arc energy is simulated annealing (SA) \cite{Kirkpatrick-etal-1983}. This method has been successfully applied on the directed and undirected feedback vertex set problems \cite{Galinier-Lemamou-Bouzidi-2013,Qin-Zhou-2014,Zhou-2016b}. For the present FAS problem we follow the simple recipe of \cite{Galinier-Lemamou-Bouzidi-2013} (Appendix~\ref{sec:appSA}). Starting from an initial random permutation and an initial low inverse temperature $\beta$, at each time step two rejection-free updating processes are performed: (1) an upward arc $(i, j)$ with $h_i < h_j$ is chosen among all such arcs with probability proportional to $\exp\bigl[-\beta \max(0, s_{(i,j)}^{i \uparrow})\bigr]$ and node $i$ is moved to be immediately above node $j$ in permutation $\mathcal{P}$, where $s_{(i,j)}^{i\uparrow}$ is the increase in the number of upward arcs caused by this move; and (2) an upward arc $(i^\prime, j^\prime)$ is chosen among all such arcs with probability proportional to $\exp\bigl[- \beta \max(0, s_{i^\prime, j^\prime}^{j^\prime \downarrow})\bigr]$ and  $j^\prime$ is moved to be immediately below node $i^\prime$ in $\mathcal{P}$, where again $s_{(i^\prime,j^\prime)}^{j^\prime \downarrow}$ is the increase in the number of upward arcs caused by this move. After $c_0 N$ such time steps (e.g., $c_0=5$ or even larger) the inverse temperature is increased to $\beta \leftarrow \beta / c_1$ (e.g., $c_1=0.99$). The search process terminates at a sufficiently large value of $\beta$.

\section{Results on random network instances}

We first test the BPD and SA algorithms on directed Erd\"os-R\'enyi (ER), directed regular random (RR) and directed scale-free (SF) random networks \cite{Dorogovtsev-Mendes-2002,Goh-Kahng-Kim-2001}.  Both ER and RR networks are homogenous, while SF networks are quite heterogeneous in that some nodes have a lot of attached arcs (Appendix~\ref{sec:appNetwork}). As the arc directions are completely random, no intrinsic flow direction should exist in these artificial networks. Our goal here is to check whether near-minimum feedback arc sets can be achieved by BPD and SA.

\begin{figure*}
  \begin{center}
    \includegraphics[angle=270,width=0.8\linewidth]{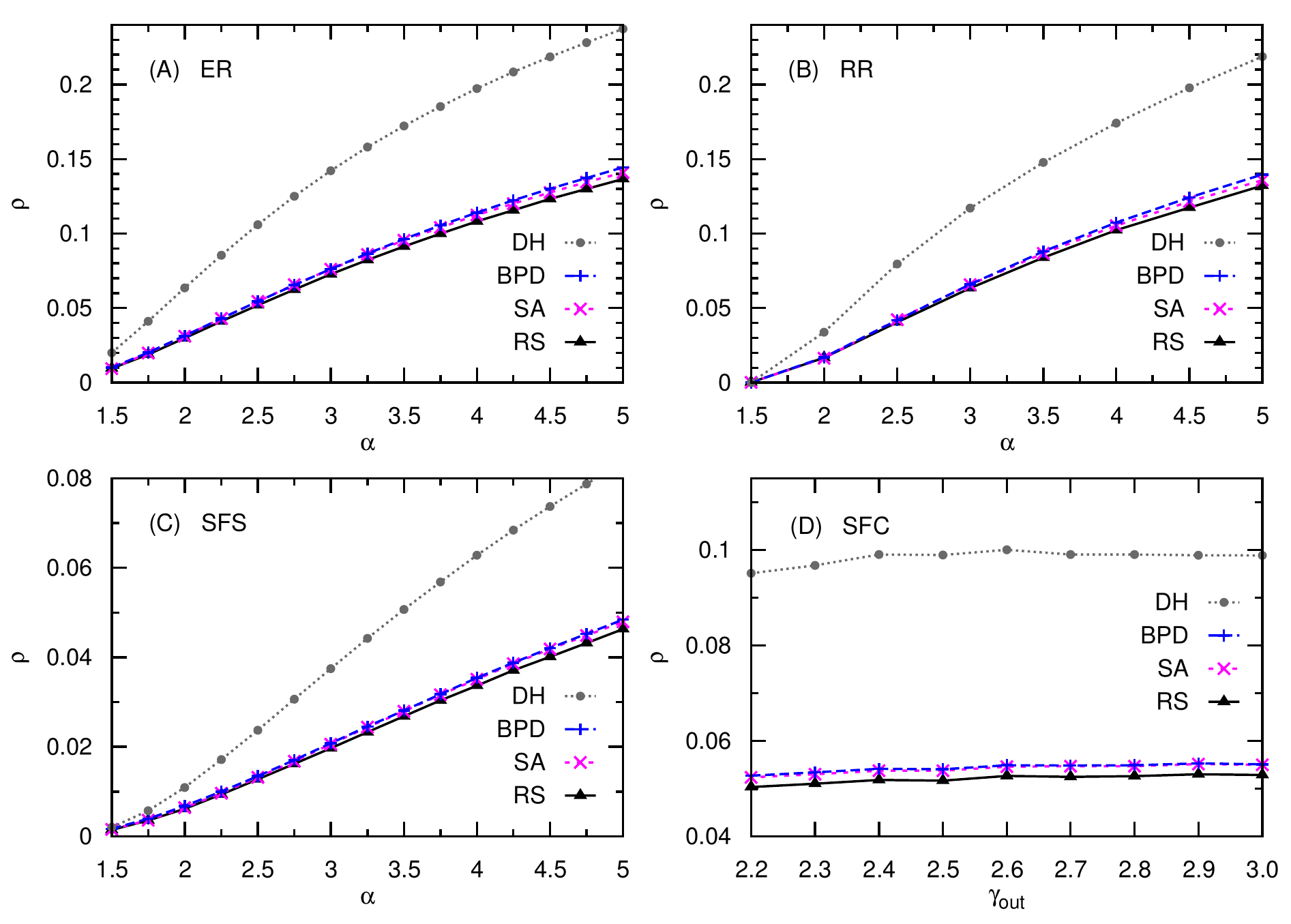}
  \end{center}
  \caption{
    \label{fig:FASrandom}
    Numerical results for random directed networks. $\alpha$, arc density; $\rho$, fraction of feedback arcs. Algorithmic results of the local DH (circles), BPD (pluses), and SA (crosses) are compared with the predictions of the RS mean-field theory (triangles).  Level upper-bound $D=200$ and inverse temperature $\beta \approx 50$ for the BPD algorithm and the RS theory. Each data point is the average over $40$ network instances of size $N=10^4$; standard deviation (not shown) is less than $4\times 10^{-3}$. Four ensembles of random networks are considered: (A) Erd\"os-R\'enyi (ER); (B) regular random (RR); (C) scale-free static (SFS \cite{Goh-Kahng-Kim-2001}) with in- and out-degree exponents $\gamma_{in}=2.5$ and $\gamma_{out}=3.0$; and (D) scale-free configurational (SFC \cite{Dorogovtsev-Mendes-2002}) with in-degree exponent $\gamma_{in}=2.5$ and different out-degree exponents $\gamma_{out}$ and minimum in- and out-degree $d_{min}=2$ and in- and out-degree upper-bound $d_{max} = \sqrt{N}$.
  }
\end{figure*}

We find that BPD and SA perform almost equally good on all the heterogeneous (SF) random networks and on the homogeneous (ER and RR) networks of arc density $\alpha < 5$; the fractions $\rho$ of feedback arcs in the constructed FAS solutions are very close to the predicted values by the RS mean-field theory, indicating that nearly optimal solutions are indeed achieved (Fig.~\ref{fig:FASrandom}). BPD and SA greatly outperform the local degree-based heuristic (DH) which recursively deletes the arc $(i, j)$ with the highest value of $d_i^{in}\times d_j^{out}$ from the network to destroy all directed cycles \cite{Pardalos-Qian-Resende-1999}, with  the in-degree $d_i^{in}$ and out-degree $d_i^{out}$ being, respectively, the number of incoming and outgoing arcs of node $i$.

The SA algorithm slightly outperforms BPD for directed ER and RR networks of arc density $\alpha \geq 5$. For ER networks of arc density $\alpha=5.0$, the typical fraction $\rho$ of feedback arcs in solutions constructed by SA has the value $\rho \approx 0.1409$, while the corresponding value for the BPD-obtained solutions is $\rho \approx 0.1445$. We can improve the performance of BPD to a small extent by choosing a larger value $D$ of level upper-bound, but the computation cost increases linearly with $D$. It appears that, to further boost the performance of the BPD algorithm and beat the SA algorithm, we need to design a better statistical physics model for the feedback arc set problem. We plan to explore this challenging issue in a future paper.

\section{Results on real-world network instances}

As a demonstration of practical applications, we now apply BPD and SA on a small set of representative real-world directed networks (Table~\ref{tab:real}):

{\bf Regulatory}. This is the epidermal growth factor receptor (EGFR) signal transduction network \cite{Oda-etal-2005,Fiedler-etal-2013}, with $N=61$ nodes and $M=112$ arcs. Each node represents a molecular species such as kinases, phosphatase, and ions; each arc represents a directed regulatory interaction between two molecular species.

{\bf Food web}. This is the Florida Bay ecosystem network \cite{Ulanowicz-etal-1998}, containing $N=128$ nodes and $M=2106$ arcs. Each node represents a species (such as bacteria, zooplankton, shrimp) or a molecular type such as particular organic carbon, and each directed arc represents transfer of biomass between two kinds of species or molecules.

{\bf Neural}. This is the neural network of the nematode \textsl{C. elegans} \cite{White-etal-1986}, containing $N=297$ nodes and $M=2359$ arcs. Each node represents a neural cell and each arc represents a directed connection between two neurons.
  
{\bf Circuit}. This is the electronic sequential logic circuit network EC-s838  \cite{Milo-etal-2002}, containing $N=512$ nodes and $M=819$ directed connections.
 
{\bf Metabolic}. This is the metabolic network of the nematode \textsl{C. elegans} \cite{Jeong-etal-2000}, with $N=1469$ nodes and $M=3447$ arcs. Each node represents a chemical molecule or an enzyme, and each arc means that a given molecule participates in a particular enzyme-catalyzed reaction or is produced by this reaction.

{\bf Wiki-Vote}. This is the network of who-votes-on-whom among the Wikipedia administrators \cite{Leskovec-Huttenlocher-Kleinberg-2010-b}, containing $N=7115$ nodes and $M=103689$ arcs.

{\bf P2P-share}. This is the Gnutella peer-to-peer file sharing network \cite{Ripeanu-etal-2002}, containing $N=62586$ nodes and $M=147892$ arcs. Each node represents a computer server and each arc represents directed file transfer between two servers.

\begin{table}
  \caption{
    \label{tab:real}
    Solving the node hierarchy problem for real-world networks. $N$, node number; $M$, arc number; $M_s$, number of simple arcs (which have no opposite counterpart); $|\Lambda|$, number of simple feedback arcs; $|\Lambda^{Rd}|$ and $\sigma^{Rd}$, expected number of simple feedback arcs in a direction-randomized network and its standard deviation;  $R$, scarcity extent of feedback arcs. Simulation results are all obtained by the SA algorithm.
  }
  \begin{center}
    \begin{tabular}{lrrrrrrr}
      \hline
      \hline
      Network & $N$  & $M$ & $M_{s}$ & $|\Lambda|$ & $|\Lambda^{Rd}|$ & $\sigma^{Rd}$ & $R$ \\
      \hline
      Regulatory & $61$ & $112$  & $108$  & $7$ & $5.4$      & $1.6$      & $-1.0$ \\
      Food web  & $128$ & $2106$ & $2044$ & $6$ & $601$      & $9$      & $63$ \\
      Neural  & $297$  & $2359$  & $1951$ & $70$ & $405$      & $10$      & $34$ \\
      Circuit  & $512$     & $819$    & $819$ & $32$      & $24.9$      & $3.9$      & $-1.8$  \\
      Metabolic   & $1469$      & $3447$  & $3383$      & $555$      & $315$      & $10$    & $-24$\\
      Wiki-Vote  & $7115$      & $103689$      & $97835$  & $3040$      & $32185$      & $74$      & $392$  \\
      P2P-share   & $62586$      & $147892$      & $147892$  & $2269$      & $13820$      & $67$      & $172$   \\
      \hline
      \hline
    \end{tabular}
  \end{center}
\end{table}
%

\begin{figure*}
  \begin{center}
    \includegraphics[angle=270,width=0.9\linewidth]{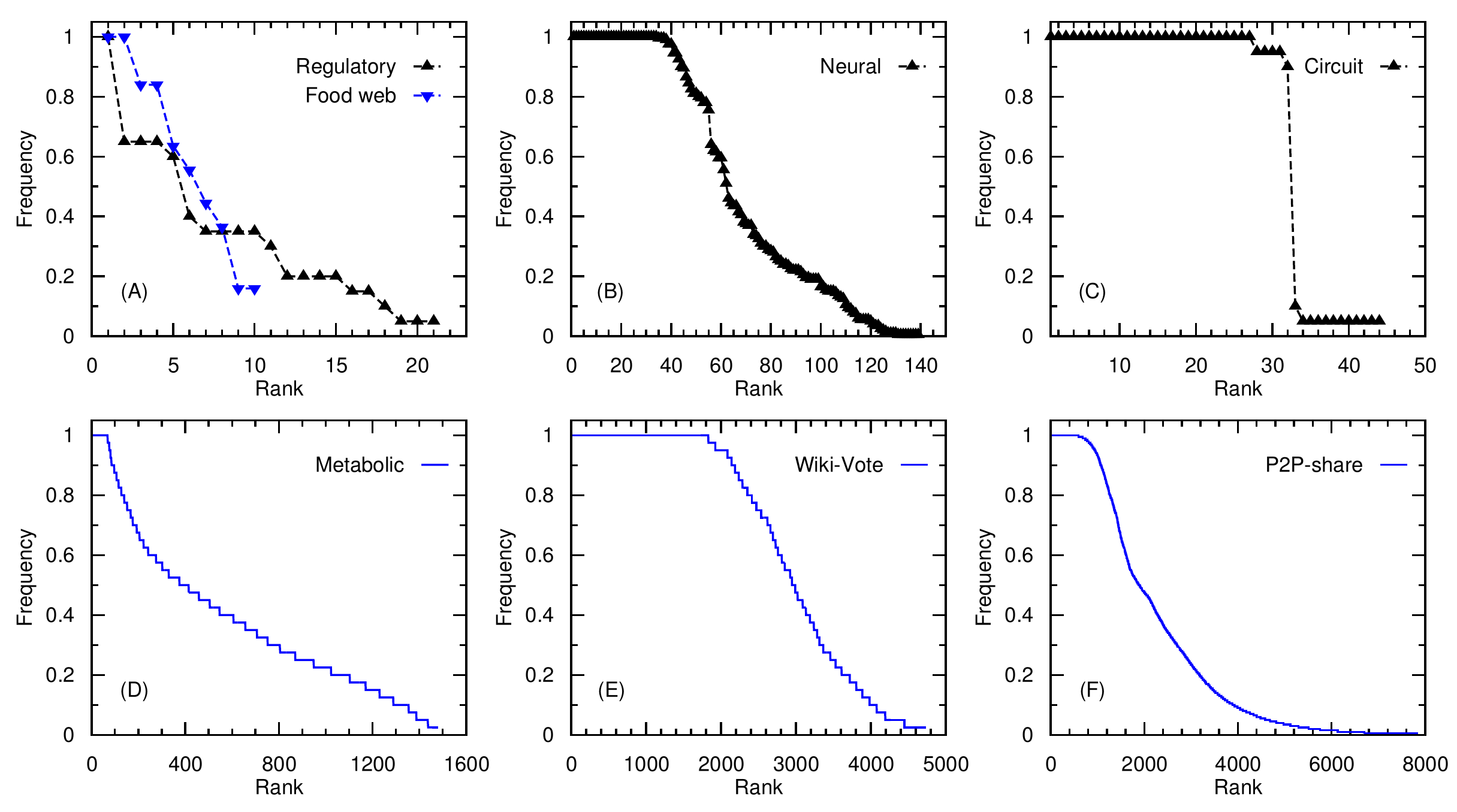}
  \end{center}
  \caption{
    \label{fig:FASreal}
    Rank plot on the frequency (probability) $\rho_{i j}$ of each arc $(i, j)$ being a feedback arc. The arcs are ranked in decreasing order according to its frequency. The results are obtained by running the SA algorithm independently for $200$ times on the same input network instance. (A) regulatory network (FAS cardinality: $7$) and food web network (FAS cardinality: $6$). (B) Neural network (FAS cardinality, mean and standard deviation: $70.0\pm 0.4$). (C) Circuit network (FAS cardinality: $32 \pm 1$). (D) Metabolic network (FAS cardinality: $556 \pm 2$). (E) Wiki-Vote network (FAS cardinality: $3038 \pm 2$). (F) P2P-share network (FAS cardinality: $2266 \pm 6$).
}
\end{figure*}

For these real-world network instances we again find that the feedback arc sets contructed by BPD and SA are of very similar sizes. It's very likely that near-optimal FAS solutions have been achieved by these two algorithms. The SA algorithm and BPD perform equally good on the four small network instances, but SA slightly outperforms BPD on the three large network instances (Metabolic, Wiki-Vote, P2P-share). We list in Table~\ref{tab:real} the results obtained by a single running of the SA algorithm on the examined real networks, where $M_s$ and $|\Lambda|$ respectively denote the total number of simple arcs and simple feedback arcs (excluding all the bi-directional arcs). 

For each examined real-world network we also generate $96$ replicas with the same connectivity pattern but completely randomized directions of all the simple arcs,  and apply SA on them to obtain the expected number $|\Lambda^{Rd}|$ of simple feedback arcs and its standard deviation $\sigma^{Rd}$. The scarcity $R$ of feedback arcs in the original network is then quantified as
\begin{equation}
  R = \frac{|\Lambda^{Rd}| - |\Lambda|}{\sigma^{Rd}} \; .
\end{equation}
This quantity has a clear statistical meaning. A large positive value of $R$ suggests that the number $|\Lambda|$ of feedback arcs in the original network is significantly lower than the expected number $|\Lambda^{Rd}|$ of feedback arcs in a direction-randomized network. Similarly, a highly negative $R$ value suggests that feedback arcs are significantly more abundant in the original network than in a direction-randomized network.

As Table~\ref{tab:real} reveals, feedback arcs are very rare in the Florida food web \cite{Ulanowicz-etal-1998}, the \textsl{C. elegans} neural network \cite{White-etal-1986}, and social networks Wiki-Vote \cite{Leskovec-Huttenlocher-Kleinberg-2010-b} and P2P-share \cite{Ripeanu-etal-2002}, which all have very large positive $R$ values. Reducing the number of feedback connections might enhance the efficiency of information processing in neural and social networks. On the other hand, feedback arcs are strongly enriched in the \textsl{C. elegans} metabolic network \cite{Jeong-etal-2000}, which has a highly negative $R$ value. It may be necessary to have an abundant number of feedback connections to finely regulate the concentrations of cellular molecules.

When we repeatedly run the SA  or BPD algorithm on the same real-world network instance, we find that the output feedback arc sets are usually not identical although their sizes are almost equal. Most importantly, we find that most of the arcs in the network never appear in any of these constructed feedback arc sets, but some arcs are present in almost all these sets (Fig.~\ref{fig:FASreal}). These results strongly indicate that the arcs in a real-world network have very different significance in terms of the feedback role, and our SA and BPD algorithms can identify a small set of most important feedback arcs. 

After the feedback probability $\rho_{i j}$ for every arc $(i, j)$ of a real-world network has been computed (through repeatedly running SA or BPD or, more efficiently, through employing Eq.~(\ref{eq:rhoarc}) and BP iteration), the feedforward part (the backbone \cite{Lan-Mezic-2011}) of the network can easily be constructed by checking every arc of the network in increasing order of the feedback probability and adding it to the backbone if no directed cycle will be formed.

\section{Conclusion and Outlook}

In this paper, we introduced the optimal node hierarchy problem, which is essentially equivalent to the minimum feedback arc set problem, and presented two physics-based algorithms to efficiently solve this problem for random and real-world directed networks. Our BPD and SA algorithms are capable of revealing the hidden hierarchical structure and the principal flow direction of a real-world directed network. Our methods can also be used to discover a small number of arcs which are involved most significantly in feedback interactions. We found that feedback interactions are extremely supressed in some real-world networks.

The methods of this work may have wide practical applications in studies of biological, technological, and social networks and in network engineering. For example, after the intrinsic flow direction in the network has been determined, it may become much more easier to design efficient arc-deletion or arc-addition strategies to improve the functionality of the network and to make it more robust against random failures or intentional attacks. The key feedback arcs identified by our algorithms may serve as optimal targets of intervening the dynamical processes on the network.

A natural extension of the present work is to consider optimal ways of cutting long directed arcs to dismantle a directed network. Similar to the proposal of optimally dismantling an undirected network \cite{Mugisha-Zhou-2016,Braunstein-etal-2016}, we may iteratively delete the arcs that are predicted to be most important for long-range feedback interactions to break the original directed network down into many small strongly connected components. Detailed numerical study on this important network optimization problem will be reported in a separate paper.

Directed cycles are large-scale structural aspects of a directed network. They cause complicated global constraints to the node hierarchy and FAS problems. Further efforts are needed to improve the theoretical models and the BPD algorithm of this paper.  Indeed the two spin glass models of the present paper still have major shortcomings. Firstly, each node $i$ of the network can take many different level states $h_i$, which considerably slows down the numerical computation. Secondly, the predicted minimum cardinalities of feedback arc sets by the two models differ noticably with each other and with the algorithmic results of BPD and SA. Thirdly, the associated BPD algorithms of the two models perform worse than the SA algorithm on homogeneous random networks of relative large arc densities. We hope these issues will be overcome in the near future by a refined statistical physics model of the minimum feedback arc set problem.

\section*{Acknowledgement}

This research was partially supported by the National Basic Research Program of China (grant number 2013CB932804) and by the National Natural Science Foundations of China (grant number 11121403 and 11225526).


\clearpage
\widetext

\begin{appendix}
  
  \begin{center}
    {\Large \bf
      Feedback arcs and node hierarchy in directed networks
    }
  \end{center}
  
  \begin{center}
    Jin-Hua Zhao and Hai-Jun Zhou
  \end{center}
  
  \vskip 0.2cm
  
  \begin{center}
    {\Large Appendices}
  \end{center}
  
  \vskip 0.5cm
  
  We describe in the following appendices the technical details of the replica-symmetric (RS) mean-field theories, the belief-propagation--guided decimation (BPD) algorithms and the simulated annealing (SA) algorithm. Some technical details on generating random directed network instances and on visualizing directed networks are also given. The source codes of the BPD and SA algorithms will be made publicly available at {\tt http://power.itp.ac.cn/$\sim$zhouhj/codes.html}.

  A directed network $G$ is composed of $N$ nodes and $M$ arcs, each of which is a directed link pointing from one node (say $i$) to another node (say $j$) and is denoted as $(i, j)$. The arc density is denoted as $\alpha \equiv \frac{M}{N}$. Given an arc $(i, j)$, we say that node $j$ is a downstream neighbor (child) of $i$ and node $i$ an upstream neighbor (parent) of $j$. Let us denote by $c(j) \equiv \{k : (j, k) \in G\}$ the set of downstream neighbors (childrens) of node $j$, and similarly denote by $p(j) \equiv \{i: (i, j)\in G\}$ the set of upstream neighbors (parents) of node $j$. The in-degree $d_j^{in}$ of node $j$ is then the cardinality of set $p(j)$, that is $d_j^{in} \equiv |p(j)|$, and the out-degree of node $j$ is $d_j^{out} \equiv |c(j)|$.

  An arc $(i, j)$ from node $i$ to node $j$ is referred to as a simple arc if (and only if) the oppositely directed arc $(j, i)$ from node $j$ to node $i$ is absent. If both $(i, j)$ and $(j, i)$ are present in the network, a trivial directed cycle involving nodes $i$ and $j$ will be formed, and one of these two arcs will be a feedforward arc and the other one will be a feedback arc. Because of this reason, in the present work we only consider simple arcs and neglect all the pairs of oppositely directed arcs. 
 
  \section{Relationship between the node hierarchy problem and the feedback arc set problem}
  \label{sec:appMap}

  A node hierarchy $\underline{h}=(h_1, h_2, \ldots, h_N)$ is a hierarchy level configuration involving all the nodes of network $G$. A node hierarchy $\underline{h}$ must satisfy the following two sets of constraints: (1) the level of each node $i$ must be a non-negative integer, namely $h_i \in \{0, 1, 2, \ldots, N\}$; (2) a node $j$ at positive level $h_j \geq 1$ must have outgoing arc(s) to node(s) at one level below (in other words,  there must be at least one arc $(j, k)$ from $j$ to a node $k$ at level $h_k=h_j-1$).
  
\begin{figure}
    \begin{center}
      \includegraphics[width=0.5\textwidth]{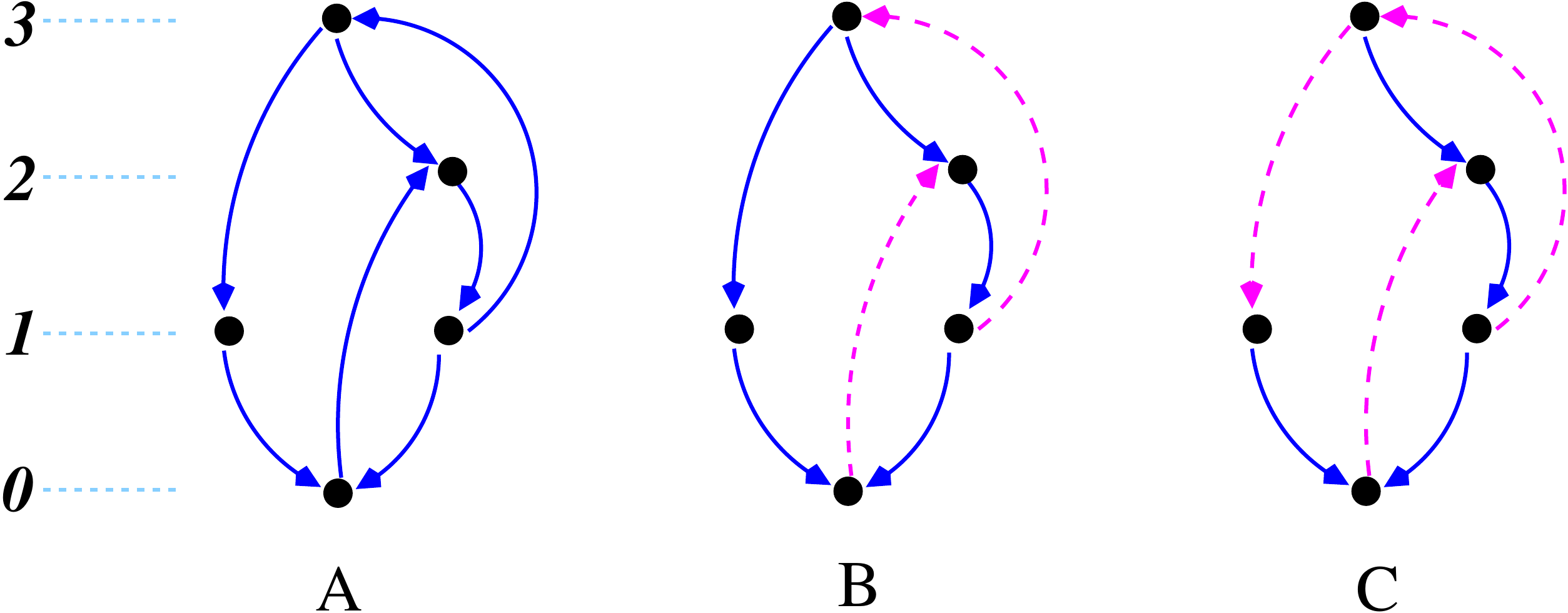}
    \end{center}
    \caption{
      \label{fig:hierarchy2fas}
      One-to-many mapping between node hierarchies and feedback arc sets. (A) a node hierarchy for a small directed network with $N=5$ nodes and $M=7$ arcs: the number of nodes at level $0$, $2$, and $3$ is one, while the number of nodes at level $1$ is two. (B) the two dashed arcs form a feedback arc set. (C) the three dashed arcs form another feedback arc set.
    }
  \end{figure}

  A feedback arc set (FAS) is a set $\Lambda$ of arcs with the property that if all the arcs in $\Lambda$ are deleted from network $G$, the remaining subnetwork will be free of any directed cycle. Two feedback arc sets for a small network are shown in Fig.~\ref{fig:hierarchy2fas} as examples. A minimal FAS is a FAS $\Lambda$ of minimal cardinality, meaning that any proper subset of $\Lambda$ is no longer a FAS. A minimum FAS is a special minimal FAS whose cardinality is the smallest among all the feedback arc sets.

  The concept of node hierarchy is closely related to the concept of feedback arc set. Given a FAS $\Lambda$, a \emph{unique} node hierarchy $\underline{h}$ can be constructed through the following iteration process:
  \begin{enumerate}
  \item[a.]
    All the arcs of set $\Lambda$ are deleted from network $G$, resulting in a directed acyclic subnetwork $G^\prime$.
  \item[b.]
    Some of the nodes must have no outgoing arc in subnetwork $G^\prime$, and they are all assigned the  lowest hierarchy level $0$. Then the level parameter $h$ is set to be $h=1$.
  \item[c.]
    If some nodes were not assigned a level during the preceding step(s), then some of these remaining nodes must  only have outgoing arcs in $G^\prime$ to the assigned nodes (whose levels are surely less than $h$), and all such nodes are assigned the hierarchy level $h$. Then $h$ is increased by one ($h \leftarrow h+1$).
  \item[d.]
    Return to step (c) as long as some nodes are still not yet assigned a hierarchy level. 
  \end{enumerate}
  The resulting hierarchy level configuration $\underline{h}=(h_1, h_2, \ldots, h_N)$ of the $N$ nodes must be a node hierarchy, as each node $i$ has integer level $h_i \geq 0$ and each node $j$ of positive level $h_j$ has at least one outgoing arc $(j, k)$ to a node $k$ of level $h_k=h_j-1$. We have therefore proved that every FAS can be mapped to a unique node hierarchy. Let us denote by $\underline{h}(\Lambda)$ the mapped node hierarchy of the FAS $\Lambda$. Because of the one-to-one mapping from $\Lambda$ to $\underline{h}(\Lambda)$, the node hierarchy $\underline{h}(\Lambda)$ contains all the information of $\Lambda$.
  
  In this work we regard a feedback arc set $\Lambda$ as \emph{neat} if, and only if, for each arc $(i, j)\in \Lambda$ the level of the upstream node $i$ does not exceed that of the downstream node $j$ (namely, $h_i \leq h_j$) in the uniquely determined node hierarchy $\underline{h}(\Lambda)$. Notice that a neat FAS of network $G$ is not necessarily a minimal FAS of $G$. (For example, the arc set containing all the arcs of the network is a neat FAS, but it is not a minimal FAS.) On the other hand, every minimal FAS must be a neat FAS. As a corollary, every minimum FAS is a neat FAS. The concept of neat FAS can be understood as a natural extension of the concept of minimal FAS.

  We now prove that there is a one-to-one correspondence between a neat FAS $\Lambda$ and a node hierarchy $\underline{h}$, which means that the $``$function'' $\underline{h}=\underline{h}(\Lambda)$ is invertible if $\Lambda$ is restricted to be a neat FAS.

  First, let us emphasize that, although there is a one-to-one mapping from FAS to node hierarchy, the mapping from node hierarchy to FAS is not one-to-one but instead many-to-one. In other words, many different feedback arc sets are compatible with the same node hierarchy $\underline{h}$.  Figure~\ref{fig:hierarchy2fas} gives a clear demonstration of this important fact. However, given a node hierarchy, say $\underline{h}^{(1)}=(h_1^{(1)}, h_2^{(1)}, \ldots, h_N^{(1)})$, a unique and neat FAS (denoted as $\Lambda^{(1)}$) can be constructed through the following simple process, starting from $\Lambda^{(1)}=\emptyset$: For each arc $(i,j) \in G$ we add it to $\Lambda^{(1)}$ if and only if $h_i^{(1)} \leq h_j^{(1)}$. It is straightforward to check that the resulting set $\Lambda^{(1)}$ must be unique, it must be a FAS, and it must be neat. If one applies on $\Lambda^{(1)}$ the above-mentioned mapping of FAS to node hierarchy, the resulting node hierarchy $\underline{h}^{(2)} \equiv \underline{h}(\Lambda^{(1)})$ will be identical to $\underline{h}^{(1)}$. This last statement can be verified by the following iterative reasoning:
  \begin{enumerate}
  \item[1.]
    If a node $i$ has level $h_i^{(1)} = 0$ in node hierarchy $\underline{h}^{(1)}$, then all its outgoing arcs must belong to the neat FAS $\Lambda^{(1)}$ and therefore it will be assigned the hierarchy level $h_i^{(2)}=0$ when $\Lambda^{(1)}$ is mapped back to a node hierarchy. Therefore $\underline{h}(\Lambda^{(1)})$ is identical to $\underline{h}^{(1)}$ at hierarchy level $0$.
  \item[2.]
    If a node $j$ has level $h_j^{(1)}=1$ in node hierarchy $\underline{h}^{(1)}$, then at least one of its outgoing arc,  say $(j, k)$, points to a node $k$ of level $h_k^{(1)}=0$ and therefore does not belong to $\Lambda^{(1)}$; on the other hand, if node $j$ has an outgoing arc, say $(j, l)$, to a node $l$ of level $h_l^{(1)} \geq 1$, this arc must belong to $\Lambda^{(1)}$. Because of these two properties, node $j$ will be  assigned the hierarchy level $h_j^{(2)}=1$ in the
    mapping $\underline{h}(\Lambda^{(1)})$. This means that $\underline{h}(\Lambda^{(1)})$ is identical to $\underline{h}^{(1)}$ also at hierarchy level $1$.
  \item[3.]
    The reasoning of the preceding step (2) can be applied to nodes at the hierarchy level $h=2, 3, \ldots$ of $\underline{h}^{(1)}$ to confirm that $\underline{h}(\Lambda^{(1)})$ is identical to $\underline{h}^{(1)}$ at all these hierarchy levels.
  \end{enumerate}

  We have therefore completed the proof of one-to-one correspondence between a node hierarchy $\underline{h}$ and a neat FAS $\Lambda$. In this work we regard a node hierarchy $\underline{h}$ as minimal if, and only if, its corresponding neat FAS $\Lambda$ is a minimal FAS. A node hierarchy $\underline{h}$ is referred to as an optimal (or minimum) node hierarchy if and only if the corresponding neat FAS $\Lambda$ is a minimum FAS.
  
  Since every minimal FAS is a neat FAS, the nice property of one-to-one correspondence between node hierarchy and neat FAS means that, the problem of constructing an optimal (or nearly optimal) node hierarchy is essentially equivalent to the problem of constructing a minimum (or nearly minimum) FAS. Because the FAS problem is a NP-complete combinatorial optimization problem, the node hierarchy problem must also be NP-complete.

  \section{Replica-symmetric mean field theory for the strongly constrained  model}
  \label{sec:appModelR}

  \begin{figure}
    \begin{center}
      \includegraphics[width=0.175\textwidth]{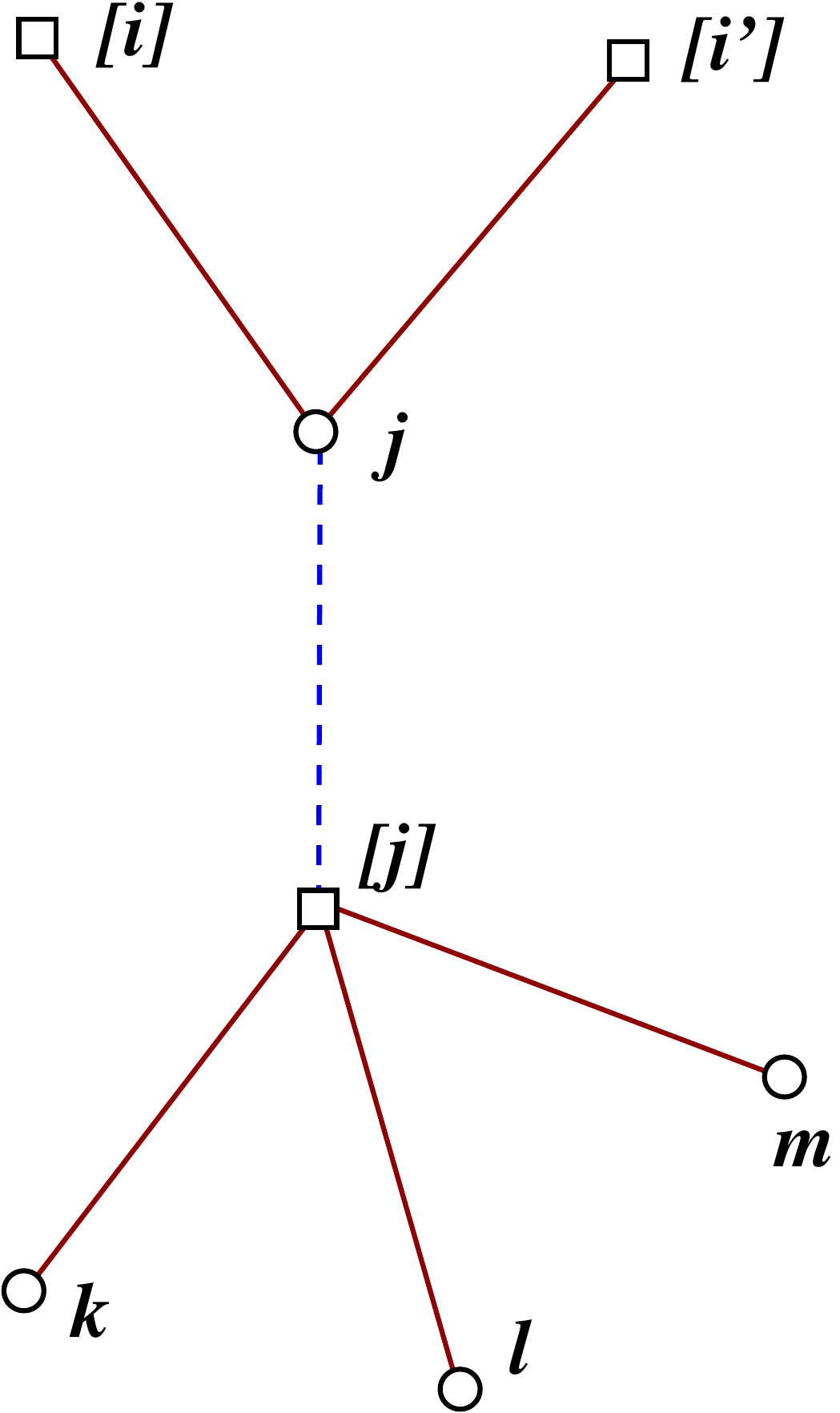}
    \end{center}
    \caption{
      \label{fig:bipartiteR}
      A bipartite-graph representation of the strongly constrained model (\ref{eq:modelRb}), showing the neighborhood structure of a node $j$. In the original directed network $G$, node $j$ has three downstream neighbors ($d_j^{out} = 3$ and $c(j) = \{k, l, m\}$) and two upstream neighbors ($d_j^{in}=2$ and $p(j)=\{i, i^\prime\}$). The circles represent node $j$ and its three downstream nodes $k$, $l$, and $m$. The squares represent the level constraints $[j], [i], [i^\prime]$ associated with nodes $j$ and all its upstream neighbors. The square for constraint $[j]$ is connected to node $j$ by a dashed line and to the  nodes in set $c(j)$ by solid lines.  
    }
  \end{figure}

  Let us refer to the strongly constrained statistical physics system (\ref{eq:modelR}) as model-R (the ``restricted'' model). According to this model, each node $j$ of the directed network $G$ has a hierarchical level $h_j \geq 0$, and the level constraint associated with node $j$ is: if $j$ is at a positive level ($h_j > 0$) then it must have at least one outgoing arc to a node at one level below (i.e., it must have an outgoing arc $(j, k)$ to a node $k$ at level $h_k=h_j-1$). The partition function of model-R at inverse temperature $\beta$ is 
  \begin{equation}
    Z_{modelR}(\beta) = \sum\limits_{h_1 = 0}^{D-1} \ldots \sum\limits_{h_N=0}^{D-1}
    \prod\limits_{i=1}^{N} \Bigl[1- (1-\delta_{0}^{h_i}) \prod\limits_{j\in c(i)}(1-
      \delta_{h_{j}+1}^{h_i}) \Bigr] \prod\limits_{(k,l)\in G} \psi_{k l}(h_k, h_l)
    \; ,
    \label{eq:modelRb}
  \end{equation}
  where the Kronecker symbol $\delta_{m}^{n} = 1$ for $m=n$ and $\delta_{m}^{n}=0$ for $m\neq n$;  the Boltzmann factor $\psi_{k l}(h_k, h_l) = 1$ for $h_k > h_l$ and $\psi_{k l}(h_k, h_l) = e^{-\beta}$ for $h_k \leq h_l$. Equation (\ref{eq:modelRb}) is identical to Eq.~(\ref{eq:modelR}) of the main text. The integer adjustable parameter $D$ is a level upper-bound introduced for computational convenience. In principle we should set  the level upper-bound $D$ to be $D=N$; but it turns out that the theoretical results are not sensitive to the precise value of $D$ (see Fig.~\ref{fig:RRRa5}), and actually too large a value of $D$ will deteriorate the performance of the associated BPD message-passing algorithm (see Fig.~\ref{fig:RRRa5BPDmodelB}).

  In the following discussions, let us denote by $[j]$ the level constraint associated with node $j$.  This constraint induces many-body interactions among $j$ and all its downstream neighbors in the set $c(j)$. We can represent model (\ref{eq:modelRb}) by a bipartite graph involving constraint nodes (squares) and variable nodes (circles) and the edges between the squares and circles, see Fig.~\ref{fig:bipartiteR}.

  \subsection{The belief-propagation equation}
  
  Let us denote by $p_{[j]\rightarrow j}^{h_j}$ the probability that node $j$ will be at level $h_j$ if it is \emph{only} constrained by the constraint  $[j]$. Similarly, for each downstream neighbor $k$ of node $j$, we denote by $p_{[j]\rightarrow k}^{h_k}$ the probability that $k$ will be at level $h_k$ if it is \emph{only} constrained by the constraint $[j]$. If node $j$ has no downstream neighbor (i.e., its out-degree $d_j^{out}=0$), then $p_{[j]\rightarrow j}^{h_j} = \delta_{0}^{h_j}$. In the general case of $d_j^{out} \geq 1$, if we assume that all the nodes attached to the constraint $[j]$ (see Fig.~\ref{fig:bipartiteR}) are mutually independent in the absence of this constraint, we can write down the following set of belief propagation (BP) equations \cite{Mezard-Montanari-2009,Mezard-Parisi-2001,Bayati-etal-2008,Altarelli-Braunstein-DallAsta-Zecchina-2013,Guggiola-Semerjian-2015,Zhou-2016b}
  \begin{subequations}
    \label{eq:BPp1}
    \begin{align}
      p_{[j]\rightarrow j}^{h_j} & =  \frac{1}{z_{[j]\rightarrow j}}
      \sum\limits_{\{h_k\,:\,k\in c(j)\}} \Bigl[
        1-(1-\delta_{0}^{h_j})\prod\limits_{k\in c(j)} 
        (1-\delta_{h_k+1}^{h_j})\Bigr]
      \prod\limits_{k\in c(j)} \Bigl[ e^{-\beta E_{j k}}
        q_{k\rightarrow [j]}^{h_k}\Bigr] \; ,  \\
      p_{[j]\rightarrow k}^{h_k} & = \frac{1}{z_{[j]\rightarrow k}}
      \sum\limits_{h_j} e^{-\beta E_{j k}} q_{j\rightarrow [j]}^{h_j}
      \sum\limits_{\{ h_{k^\prime}\,:\, k^\prime \in c(j)\backslash k\}}
      \Bigl[1-(1-\delta_{0}^{h_j}) \prod\limits_{k^\prime \in c(j)}
        (1-\delta_{h_{k^\prime}+1}^{h_j})\Bigr]
      \prod\limits_{k^\prime \in c(j)\backslash k} \Bigl[e^{-\beta E_{j k^\prime}}
        q_{k^\prime \rightarrow [j]}^{h_{k^\prime}}\Bigr] \; ,
    \end{align}
  \end{subequations}
  where $z_{[k]\rightarrow k}$ and $z_{[k]\rightarrow j}$ are two probability normalization constants; $c(j)\backslash k$ means the subset of $c(j)$ with node $k$ being excluded; $E_{j k}$ is the energy of arc $(j, k)$ which is $E_{j k}=0$ for $h_j>h_k$ and $E_{j k}=1$ for  $h_j \leq h_k$; and 
  \begin{equation}
    q_{j\rightarrow [j]}^{h_j} \equiv  
    \prod\limits_{i\in p(j)} p_{[i]\rightarrow j}^{h_j}
    \; ,
    \quad \quad 
    q_{k\rightarrow [j]}^{h_k}  \equiv 
    p_{[k]\rightarrow k}^{h_k} \prod\limits_{ j^\prime \in p(k)\backslash j}
    p_{[j^\prime]\rightarrow k}^{h_k}
    \; ,
  \end{equation}
  with $p(k)\backslash j$ being the subset of $p(k)$ with node $j$ being excluded. The quantity $q_{j\rightarrow [j]}^{h_j}$ actually is proportional to the probability that node $j$ will be at level $h_j$ if it is \emph{not} constrained by the constraint $[j]$; similarly, $q_{k\rightarrow [j]}^{h_k}$ is proportional to the probability that node $k$ will be at level $h_k$ if it is not constrained by the constraint $[j]$.
  
  The BP equation (\ref{eq:BPp1}) can be rewritten in the following equivalent form which is more convenient for numerical implementation:
  \begin{subequations}
    \begin{align}
      p_{[j]\rightarrow j}^{h_j} & \propto 
      \prod\limits_{k\in c(i)}  \Bigl[
        \sum\limits_{h_k} e^{-\beta E_{j k}} 
        q_{k\rightarrow [j]}^{h_k} \Bigr]
      - (1-\delta_{0}^{h_j}) \prod\limits_{k\in c(j)} \Bigl[
        \sum\limits_{h_k}  e^{-\beta E_{j k}}
        q_{k\rightarrow [j]}^{h_k}
        - q_{k\rightarrow [j]}^{h_j-1} \Bigr]
      \; , 
      \\
      p_{[j]\rightarrow k}^{h_k} & \propto \sum\limits_{h_j} e^{-\beta E_{j k}}
      q_{j\rightarrow [j]}^{h_j} \biggl\{ \prod\limits_{k^\prime \in c(j)\backslash k}
      \Bigl[\sum\limits_{h_{k^\prime}} e^{-\beta E_{j k^\prime}}
        q_{k^\prime \rightarrow [j]}^{h_{k^\prime}} \Bigr] -(1-\delta_{0}^{h_j})
      (1-\delta_{h_k+1}^{h_j}) \prod\limits_{k^\prime \in c(j)\backslash k}
      \Bigl[ \sum\limits_{h_{k^\prime}} e^{-\beta E_{j k^\prime}}
        q_{k^\prime\rightarrow [j]}^{h_{k^\prime}} - q_{k^\prime \rightarrow [j]}^{h_j-1}\Bigr]
      \biggr\}\; .
    \end{align}
  \end{subequations}

  \subsection{Thermodynamic quantities}
  
  The probability $\rho_{j k}$ of arc $(j,k)$ being a feedback arc is equal to the probability of $h_j \leq h_k$. According to the RS mean-field theory (i.e., assuming that all the attached nodes of the constraint $[j]$ in Fig.~\ref{fig:bipartiteR} are mutually independent in the absence of this constraint), we have
  \begin{equation}
    \rho_{j k} = \frac{e^{-\beta}}{z_{j}} 
    \sum\limits_{h_j \geq 0}  q_{j\rightarrow [j]}^{h_j}
    \Bigl[\sum\limits_{h_k \geq h_j}  q_{k\rightarrow [j]}^{h_k}   \Bigr]
    \biggl\{  \prod\limits_{k^\prime \in c(j)\backslash k} 
    \Bigl[ \sum\limits_{h_{k^\prime}}
      e^{-\beta E_{j k^\prime}} q_{k^\prime \rightarrow [j]}^{h_{k^\prime}} \Bigr]
    - (1-\delta_{0}^{h_j}) \prod\limits_{k^\prime \in c(j)\backslash k}
    \Bigl[ \sum\limits_{h_{k^\prime}} e^{-\beta E_{j k^\prime}}
      q_{k^\prime\rightarrow [j]}^{h_{k^\prime}} -q_{k^\prime \rightarrow [j]}^{h_j-1}\Bigr]
    \biggr\} \; ,
  \end{equation}
  where $z_{j}$ is expressed as
  \begin{equation}
    z_{j} =
    \sum\limits_{h_j \geq 0}
    q_{j\rightarrow [j]}^{h_j}
    \biggl\{
    \prod\limits_{k^\prime \in c(j)} \Bigl[
      \sum\limits_{h_{k^\prime}}
      e^{-\beta E_{j k^\prime}} q_{k^\prime \rightarrow [j]}^{h_{k^\prime}} \Bigr]
    - (1-\delta_{0}^{h_j})
    \prod\limits_{k^\prime \in c(j)}
    \Bigl[ \sum\limits_{h_{k^\prime}} e^{-\beta E_{j k^\prime}}
      q_{k^\prime\rightarrow [j]}^{h_{k^\prime}}
      -q_{k^\prime \rightarrow [j]}^{h_j-1}\Bigr]
    \biggr\} \; .
  \end{equation}
  The mean fraction $\rho_{modelR}$ of feedback arcs is then obtained through
  \begin{equation}
    \rho_{modelR} = \frac{1}{M} \sum\limits_{(j,k)\in G} \rho_{j k} \; .
  \end{equation}

  The free energy $F_{modelR}(\beta) \equiv -\frac{1}{\beta}\ln Z_{modelR}(\beta)$ of the whole system is computed through  \cite{Mezard-Montanari-2009,Mezard-Parisi-2001,Bayati-etal-2008,Altarelli-Braunstein-DallAsta-Zecchina-2013,Guggiola-Semerjian-2015,Zhou-2016b}
  \begin{equation}
    \label{eq:FmodelR}
    F_{modelR}(\beta)  = \sum\limits_{j=1}^{N} \Bigl[f_{[j]}- d_{j}^{in} f_{j} \Bigr]
    \; ,
  \end{equation}
  where $f_{[j]}$ is the free energy contribution of constraint $[j]$, and $f_j$ is the free energy contribution of node $j$. The expressions for these two free energy contributions are, respectively,
  \begin{subequations}
    \begin{align}
      f_{j} & = -\frac{1}{\beta} \ln \biggl\{\sum\limits_{h_j}
      p_{[j]\rightarrow j}^{h_j} 
      q_{j\rightarrow [j]}^{h_j} \biggr\}
      \; ,
      \\
      f_{[j]} & = -\frac{1}{\beta} \ln \biggl\{
      \sum\limits_{h_j} q_{j\rightarrow [j]}^{h_j}
      \sum\limits_{\{h_k\,:\,k\in c(j)\}}
      \Bigl[1- (1-\delta_0^{h_j}) \prod\limits_{k\in c(j)}
        (1-\delta_{h_k +1}^{h_j}) \Bigr] 
      \prod\limits_{k\in c(j)} \Bigl[e^{-\beta E_{j k}}
        q_{k\rightarrow [j]}^{h_k} \Bigr] \biggr\} \nonumber \\
      & = -\frac{1}{\beta} \ln \biggl\{
      \sum\limits_{h_j} q_{j\rightarrow [j]}^{h_j}
      \Bigl[\prod\limits_{k\in c(j)} 
        \bigl[\sum\limits_{h_k} e^{-\beta E_{j k}} 
          q_{k\rightarrow [j]}^{h_k} \bigr]
        - (1-\delta_{0}^{h_j}) \prod\limits_{k\in c(j)} \bigl[
          \sum\limits_{h_k}  e^{-\beta E_{j k}}
          q_{k\rightarrow [j]}^{h_k}
          - q_{k\rightarrow [j]}^{h_j-1} \bigr]
        \Bigr]
      \biggr\}
      \; .
    \end{align}
  \end{subequations}
  The free energy density is then $f_{modelR} \equiv \frac{1}{N} F_{modelR}(\beta)$. Notice that the free energy contribution $f_{[j]}$ of a constraint $[j]$ also contains the contributions of node $j$ and all its downstream neighbors (see Fig.~\ref{fig:bipartiteR}), therefore the free energy contribution $f_j$ of a node $j$ has been considered $(d_j^{in}+1)$ times in the first summation of Eq.~(\ref{eq:FmodelR}). The second summation of Eq.~(\ref{eq:FmodelR}) corrects this over-counting.
    
    The entropy density $s_{modelR}$ of the system is then
  \begin{equation}
    s_{modelR} = \beta \bigl( \alpha \rho_{modelR}  - f_{modelR} \bigr) \; .
  \end{equation}
  This expression can be understood from the relationship that $Z_{modelR} \approx \exp\bigl(- \beta M \rho_{modelR} + N s_{modelR} \bigr)$.
  
  \subsection{Theoretical and algorithmic results on a special kind of regular random directed networks}

  \begin{figure}[t]
    \begin{center}
      \includegraphics[angle=270,width=0.5\textwidth]{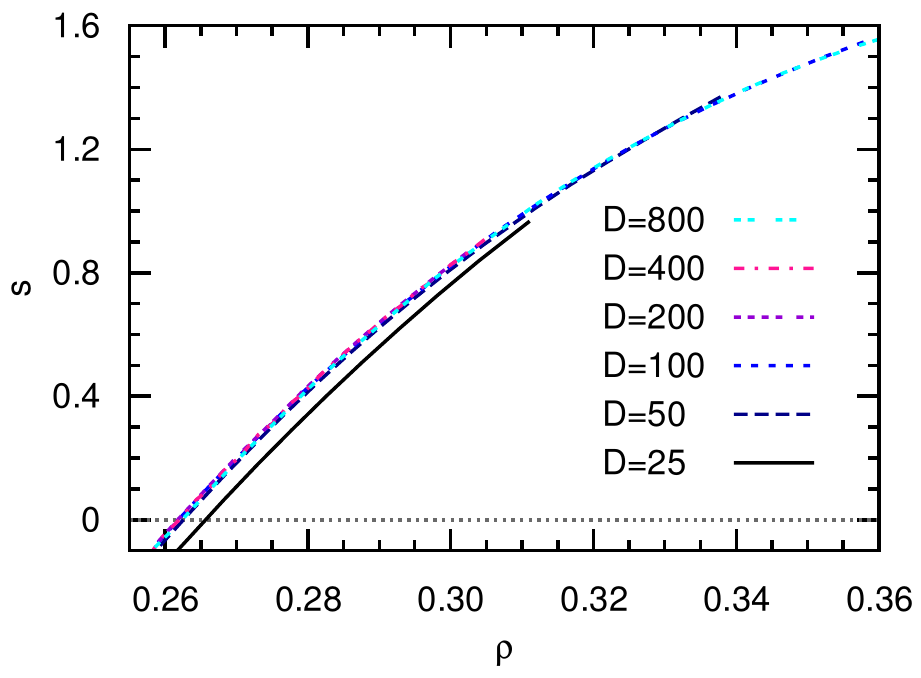}
    \end{center}
    \caption{
      \label{fig:RRRa5}
      Replica-symmetric mean field results (based on model-R) for a special type of regular random directed networks. Each node in the random network has $K=5$ incoming arcs and $K=5$ outgoing arcs. The level upper-bound $D$ ranges from $D=25$ to $D=800$ in the mean-field computations. $s$, entropy density; $\rho$, fraction of feedback arcs.
    }
  \end{figure}

  Due to the many-body nature of the node level constraints, the above-mentioned RS mean field theory is computationally quite inefficient. As a first test of this theory, we apply it to a special kind of regular random directed networks, namely random directed networks in which each node has the same number $K$ of incoming arcs and the same number $K$ of outgoing arcs. The connectivity pattern of such a random network is otherwise completely random. Such a network may be referred to as a balanced random regular (BRR) network.  For this BRR network ensemble we can assume that the cavity probability distributions $p_{[j]\rightarrow j}^{h}$ are independent of the node $j$ and are all equal to the same distribution $p_{con-to-self}^{h}$, and similarly all the cavity probability distributions $p_{[j]\rightarrow k}^{h}$ are independent of the node $j$ and the downstream neighbor $k$ but are equal to the same distribution $p_{con-to-down}^{h}$. Under these two additional assumptions the BP equation (\ref{eq:BPp1}) can be simplified and a fixed-point solution can be obtained by numerical iterations.

 Figure~\ref{fig:RRRa5} shows the theoretically predicted relationship between entropy density and feedback arc fraction at $K=5$. As long as the level upper-bound $D\geq 50$, the RS theoretical results are almost independent of $D$. At feedback arc fraction $\rho=0.2620$ the entropy density changes from being positive to being negative. Therefore this mean field theory predicts  the minimum fraction of feedback arcs to be $\rho = 0.2620$.

  \begin{figure}
    \begin{center}
      \includegraphics[angle=270,width=0.5\textwidth]{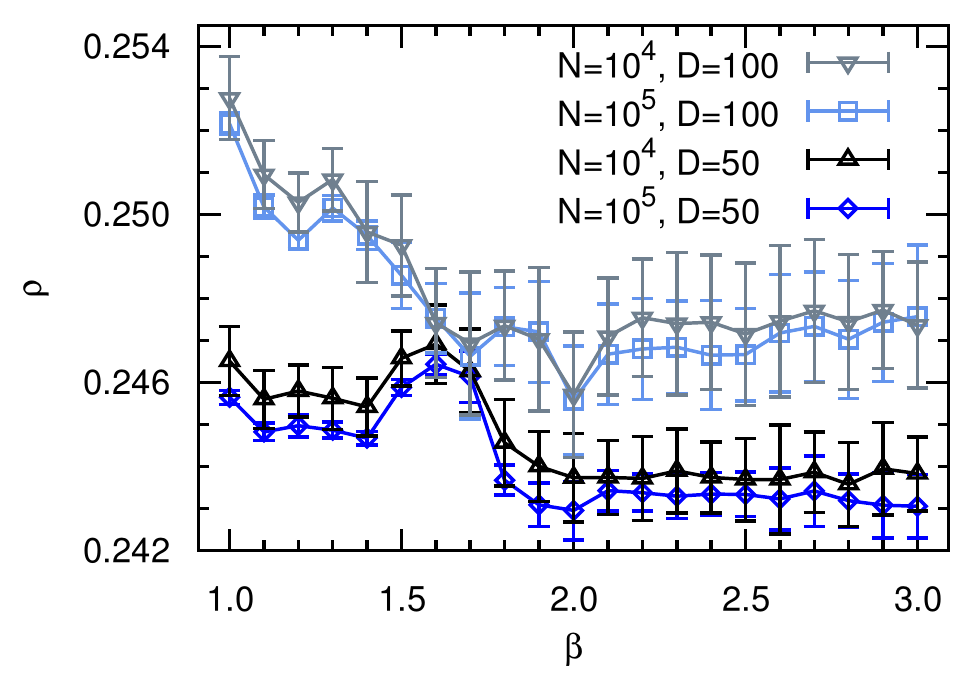}
    \end{center}
    \caption{
      \label{fig:RRRa5BPDmodelB}
      Results obtained by the BPD algorithm (based on model-R) for two balanced regular random directed network instances (one of size $N=10^4$ and the other of size $N=10^5$), in which each node has $K=5$ incoming arcs and $K=5$ outgoing arcs. Each data point is averaged over $48$ independent runs of the BPD algorithm (the level upper-bound is fixed to $D=50$ or $D=100$). $\rho$, fraction of feedback arcs in the constructed solutions; $\beta$, the inverse temperature of the BPD algorithm.
    }
  \end{figure}

  We can implement a BPD message-passing algorithm based on the BP equation (\ref{eq:BPp1}) to construct feedback arc sets for single  BRR network instances (see Appendix~\ref{sec:appBPD}). We have applied this BPD algorithm to two single network instances, one with $N=10^4$ nodes and the other with $N=10^5$ nodes ($K=5$ in both cases). Figure~\ref{fig:RRRa5BPDmodelB} shows the averaged results over $48$ independent runs of the BPD algorithm at various different values of the inverse temperature $\beta$. We notice that the BPD results obtained at $D=50$ are slightly better than those obtained at $D=100$; we also notice that the BPD algorithm is not sensitive to the value of the inverse temperature $\beta$ as long as $\beta\geq 2$.  Most strikingly, we find that the mean fraction $\rho\approx 0.243$ of feedback arcs in solutions obtained by the BPD algorithm is noticeably less than the predicted minimum fraction $0.262$ by the RS mean field theory of model-R. Indeed for the studied network instance of size $N=10^5$, the feedback arc fraction $\rho$ in the best solution obtained by BPD is $\rho=0.2417$. The associated BPD algorithm of the relaxed model (model-E, see Appendix~\ref{sec:appModelE}) gives very close results ($\rho \approx 0.2410$) for this network instance, while the SA algorithm (see Appendix~\ref{sec:appSA}) achieves even better solutions with $\rho \approx 0.2212$.

  The fact that the theoretically predicted minimum feedback arc fraction being higher than the results obtained by the BPD and SA algorithms indicates that the level constraints of model-R are too strong for the feedback arc set problem.

  \section{The replica-symmetric mean-field theory for the relaxed model}
  \label{sec:appModelE}
  
  Let us refer to the relaxed model as model-E (the ``easy'' model). There is essentially no constraint in this model except that the hierarchical level $h_i$ of each node $i$ should be an integer in the range $[0, D-1]$. Each directed arc $(j, k)$ causes a two-body interaction between node $j$ and node $k$, and the partition function of model-E is
  \begin{equation}
    Z_{modelE}(\beta) = \sum\limits_{h_1=0}^{D-1} \ldots
    \sum\limits_{h_N=0}^{D-1} \prod\limits_{(j, k)\in G}
    \psi_{j k}(h_j, h_k) \; ,
    \label{eq:modelE}
  \end{equation}
  where $\psi_{j k}(h_j, h_k) = 1$ for $h_j > h_k$ and $\psi_{j k}(h_j, h_k)= e^{-\beta}$ for $h_j \leq h_k$. The RS mean-field theory for this partition function has already been briefly described in the main text, here we add some more technical explanations.
  
  \subsection{The belief-propagation equation}
  
  For an arc $(i,j)$ from node $i$ to node $j$, its probability of being a feedback arc is
  \begin{equation}
    \label{eq:rhoijE}
    \rho_{i j}=\frac{
      e^{-\beta}\sum\limits_{h_i= 0}^{D-1} q_{i\rightarrow j}^{h_i}
      \sum\limits_{h_j=h_i}^{D-1} q_{j\rightarrow i}^{h_j}}
        { 1-(1-e^{-\beta}) \sum\limits_{h_i = 0}^{D-1} q_{i\rightarrow j}^{h_i}
          \sum\limits_{h_j=h_i}^{D-1} q_{j\rightarrow i}^{h_j}}
        \; ,
  \end{equation}
  where $q_{i\rightarrow j}^{h_i}$ denotes the cavity probability that node $i$ will be at level $h_i$ if node $j$ is absent; similarly $q_{j\rightarrow i}^{h_j}$ is the cavity probability that node $j$ will be at level $h_j$ if node $i$ is absent. Notice that in Eq.~(\ref{eq:rhoijE}) the product $q_{i\rightarrow j}^{h_i} q_{j\rightarrow i}^{h_j}$ is the joint probability of $h_i$ and $h_j$ when the arc $(i, j)$ is absent (assuming that node $i$ and node $j$ are then independent); and the term $\sum_{h_i \geq 0}\sum_{h_j \geq h_i} q_{i\rightarrow j}^{h_i} q_{j\rightarrow i}^{h_j}$ is then the total probability that $h_i \leq h_j$ in the absence of the arc $(i, j)$. The belief-propagation equation for the cavity probabilities are expressed as  \cite{Mezard-Montanari-2009,Mezard-Parisi-2001,Bayati-etal-2008,Altarelli-Braunstein-DallAsta-Zecchina-2013,Guggiola-Semerjian-2015,Zhou-2016b}
  \begin{subequations}
    \label{eq:BPmodelE}
    \begin{align}
      q_{j \rightarrow k}^{h_j}  & = \frac{1}{z_{j\rightarrow k}}
      \prod\limits_{i\in p(j)} \Bigl[e^{- \beta} + (1- e^{-\beta})
        \sum\limits_{h_i = h_j+1}^{D-1} q_{i\rightarrow j}^{h_i} \Bigr] 
      \prod\limits_{k^\prime \in c(j)\backslash k} \Bigl[e^{-\beta} + (1-e^{-\beta}) 
        \sum\limits_{h_{k^\prime} = 0}^{h_j} q_{k^\prime \rightarrow j}^{h_{k^\prime}} 
        \Bigr] \; ,
      \label{eq:BPmodelE:a} \\
      q_{j \rightarrow i}^{h_j} & =  \frac{1}{z_{j\rightarrow i}}
      \prod\limits_{i^\prime \in p(j)\backslash i} \Bigl[
        e^{- \beta} + (1- e^{- \beta}) \sum\limits_{h_{i^\prime}= h_j+1}^{D-1}
        q_{i^\prime \rightarrow j}^{h_{i^\prime}} \Bigr]  \prod\limits_{k \in c(j)}
      \Bigl[e^{- \beta}+(1-e^{-\beta})\sum\limits_{h_k = 0}^{h_j}
        q_{k \rightarrow j}^{h_k}
        \Bigr] \; , \label{eq:BPmodelE:b}
    \end{align}
  \end{subequations}
  where node $k$ in Eq.~(\ref{eq:BPmodelE:a}) belongs to set $c(j)$ and node $i$ in Eq.~(\ref{eq:BPmodelE:b}) belongs to set $p(j)$; $z_{j\rightarrow k}$ and $z_{j\rightarrow i}$ are two probability normalization constants. Notice that Eq.~(\ref{eq:BPmodelE}) is equivalent to Eq.~(\ref{eq:BP}) of the main text.
  
  \subsection{Thermodynamic quantities}
  
  At a given value of the inverse temperature $\beta$, we can compute the mean fraction of feedback arcs as
  \begin{equation}
    \rho_{modelE}  = \frac{1}{M} \sum\limits_{(i,j)\in G} \rho_{i j} \; .
  \end{equation}
  The total free energy of model-E, defined by $F_{modelE}= - \frac{1}{\beta} \ln Z_{modelE}(\beta)$, can be expressed as  \cite{Mezard-Montanari-2009,Mezard-Parisi-2001,Bayati-etal-2008,Altarelli-Braunstein-DallAsta-Zecchina-2013,Guggiola-Semerjian-2015,Zhou-2016b}
  \begin{equation}
    \label{eq:FmodelE}
    F_{modelE} = \sum\limits_{j=1}^{N} f_j - \sum\limits_{(k, l) \in G} f_{k l} \; ,
  \end{equation}
  where $f_j$ and $f_{k l}$ are, respectively, the node and arc contribution to the free energy:
  \begin{subequations}
    \label{eq:fvandarc}
    \begin{align}
      f_j & = -\frac{1}{\beta} \ln \biggl\{ \sum\limits_{h_j=0}^{D-1}
      \prod\limits_{i\in p(j)} \Bigl[1 - (1- e^{- \beta}) \sum\limits_{h_i=0}^{h_j}
        q_{i\rightarrow j}^{h_i} \Bigr] \prod\limits_{k \in c(j)}
      \Bigl[1-(1-e^{-\beta}) \sum\limits_{h_k=h_j}^{D-1}  q_{k \rightarrow j}^{h_k}
        \Bigr] \biggr\} \; , 
      \\
    f_{k l} & = -\frac{1}{\beta} \ln \biggl\{ 1- (1 - e^{-\beta})
    \sum\limits_{h_k =0}^{D-1} \sum\limits_{h_l=h_k}^{D-1} q_{k\rightarrow l}^{h_k}
    q_{l\rightarrow k}^{h_l} \biggr\} \; .
    \end{align}
  \end{subequations}
  To understand Eq.~(\ref{eq:FmodelE}) in an intuitive way, we notice that the free energy contribution $f_j$ of each node $j$ includes the contributions from all the attached outgoing and incoming arcs, therefore the contribution of an arc $(j, k)$ is considered twice (in $f_j$ and $f_k$); such an over-counting is corrected by the second summation of Eq.~(\ref{eq:FmodelE}).
  
  The free energy density is simply $f_{modelE} \equiv \frac{1}{N} F_{modelE}$. And the entropy density $s_{modelE}$ at a given value of $\beta$ is then evaluated as
  \begin{equation}
    \label{eq:entropyE}
    s_{modelE} = \beta \bigl[ \alpha \rho_{modelE} - f_{modelE} \bigr] \; ,
  \end{equation}
  where $\alpha$ is the arc density. Equation (\ref{eq:entropyE}) is justified by the fact that $Z_{modelB} \approx \exp\bigl( - M \beta \rho_{modelE} + N s_{modelE}\bigr)$. 
  
  \subsection{Computation for single network instances and for an ensemble of networks}
  
  The RS mean-field theory can be applied on single instances of directed networks. Given a directed network $G$, we first iterate the BP equation (\ref{eq:BPmodelE}) on all the directed arcs a number $t_0$ (e.g., $t_0=200$) of times to reach a fixed point or to bring the set of all the cavity probability distributions $\{q_{j\rightarrow i}^{h_j}, q_{j\rightarrow k}^{h_j}\}$ close to a steady state. Then we repeat the BP iteration an additional number $t_1$ (e.g., $t_1=1000$) of times, at each time step we compute all the node free energy contributions $f_j$, all the arc free energy contributions $f_{k l}$ and probabilities $\rho_{k l}$ to evaluate the values of $f_{modelE}$, $\rho_{modelE}$ and $s_{modelE}$. The averaged results of $f_{modelE}$, $\rho_{modelE}$, and $s_{modelE}$ over these $t_1$ iterations are then reported as the free energy density, the mean fraction of feedback arcs, and the entropy density, respectively.

\begin{figure}
    \begin{center}
      \includegraphics[angle=270,width = 0.7\linewidth]{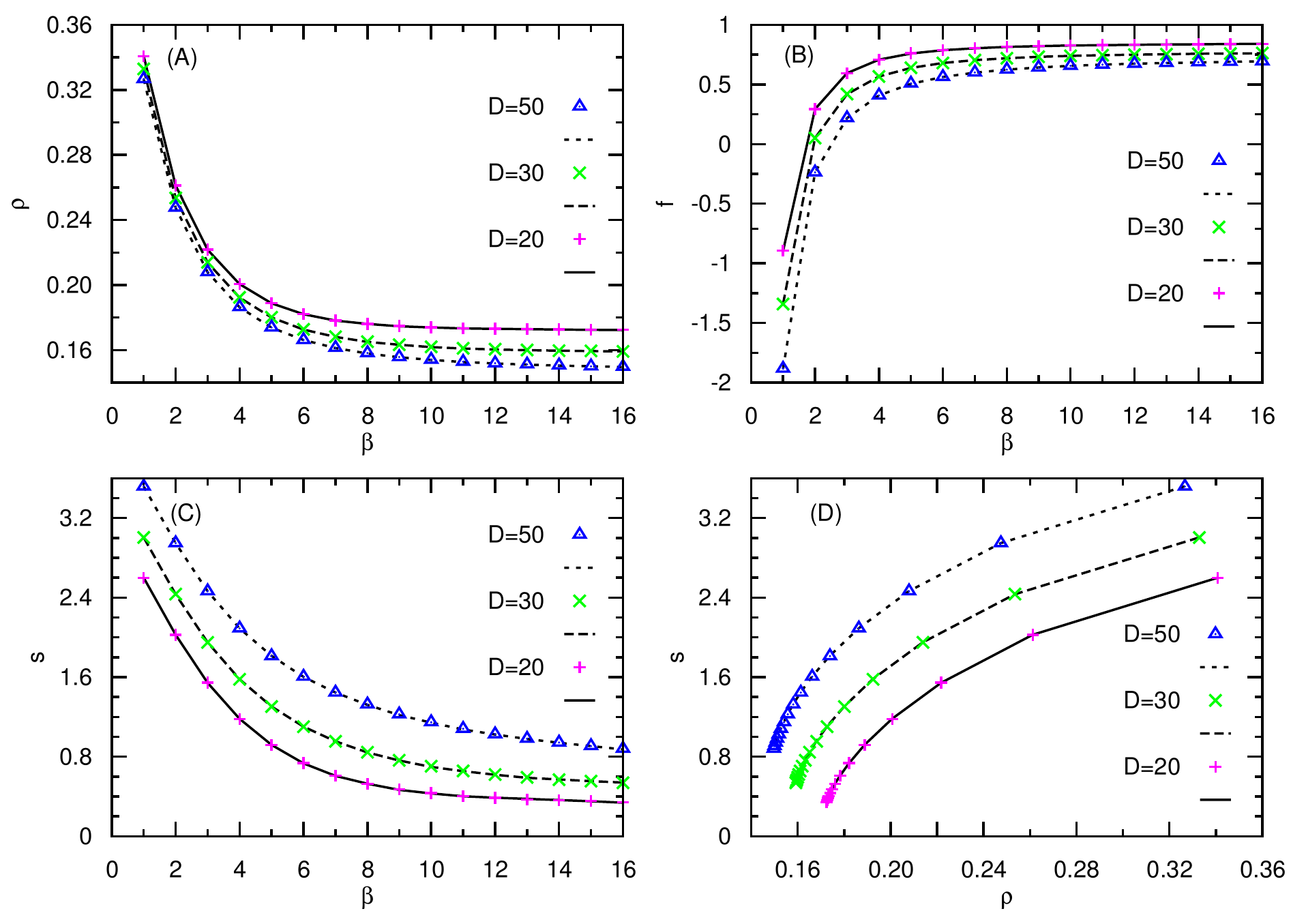}
    \end{center}
    \caption{
      \label{fig:fas_er_bp}
      Some RS mean-field theoretical results obtained on model-E (\ref{eq:modelE}) for ER random directed networks of arc density $\alpha=5.0$. The dots are the averaged simulation results obtained by BP iterations on $40$ independent ER network instances of $N=10^5$, with the level upper-bound being $D=20$ (pluses), $D=30$ (crosses), and $D=50$ (triangles. The corresponding lines are the results obtained by RS population dynamics (which corresponds to network size $N=\infty$). (A) the fraction $\rho$ of feedback arcs as a function of inverse temperature $\beta$; (B) the free energy density $f$ as a function of $\beta$; (C) the entropy density $s$ as a function of $\beta$; (D) the entropy density $s$ as a function of $\rho$, obtained by eleminating $\beta$ from data in (A) and (C).
    }
  \end{figure}

  Some BP simulation results obtained on Erd\"os-R\'enyi (ER) random directed networks of arc density $\alpha=5.0$ are shown Fig.~\ref{fig:fas_er_bp}. 

  We can also obtain ensemble-averaged results for random directed networks which have no any correlation in the connectivity pattern. For such networks, let us denote by $P(d^{in}, d^{out})$ the joint probability that a randomly chosen node has $d^{in}$ incoming arcs and $d^{out}$ outgoing arcs. We construct two large sets $\mathcal{M}_{p-to-c}$ and $\mathcal{M}_{c-to-p}$, with $\mathcal{M}_{p-to-c}$ storing many (e.g., $L=10^4$) cavity probability distributions $q_{i\rightarrow j}^{h_i}$ and $\mathcal{M}_{c-to-p}$ storing many cavity probability distributions $q_{j\rightarrow i}^{h_j}$ for different arcs $(i, j)$. We then update these two sets a large number of times (e.g., $10^4 \times L$) to drive these two sets to a steady state and to evaluate the ensemble-averaged values of the thermodynamic quantities. At each updating process, two non-negative integers $d^{in}$ and $d^{out}$ are drawn from the joint distribution $P(d^{in}, d^{out})$ and assigned to a node (say $j$) as its in-degree and out-degree, respectively. The $d^{in}$ cavity probability distributions $q_{i\rightarrow j}^{h_i}$ from the incoming arcs $(i, j)$ are then drawn from the set $\mathcal{M}_{p-to-c}$ uniformly at random and with replacement; similarly the $d^{out}$ cavity probability distributions $q_{k\rightarrow j}^{h_k}$ from the outgoing arcs $(j, k)$ are drawn from the set $\mathcal{M}_{c-to-p}$ uniformly at random and with replacement. Then $d^{in}$ new cavity probability distributions $q_{j\rightarrow i}^{h_j}$ and $d^{out}$ new cavity probability distributions $q_{j\rightarrow k}^{h_j}$ are computed according to Eq.~(\ref{eq:BPmodelE}), and they replace $d^{in}$ randomly chosen elements of set $\mathcal{M}_{c-to-p}$ and $d^{out}$ randomly chosen elements of set $\mathcal{M}_{p-to-c}$, respectively. The values of $f_{j}$ (for node $j$) and $f_{j k}$ and $f_{i j}$ and $\rho_{j k}$ and $\rho_{i j}$ for all the attached arcs of node $j$ are computed during this process. 

  As demonstrated in Fig.~\ref{fig:fas_er_bp}, the results obtained by this RS population dynamics simulation are in complete agreement with the BP results obtained on single network instances.
  
  \section{Minimizing the number of feedback arcs by belief-propagation--guided  decimation}
  \label{sec:appBPD}
  
  We have implemented two versions of the BPD algorithm based on the strongly constrained model-R (Appendix~\ref{sec:appModelR}) and  on the relaxed model-E (Appendix~\ref{sec:appModelE}), respectively. These two versions of the BPD algorithm have the same algorithmic design and the same overall structure except for the differences in the adopted BP equations. Here we describe some technical details of the algorithm. 

  Given an input directed network $G$, we first simplify it by recursively removing all the nodes which have no outgoing or incoming arcs. The arcs attached to these removed nodes are also deleted from the network. Then every node in the remaining subnetwork must have both incoming and outgoing arcs. The BPD search process is then applied on this subnetwork.
  
  Starting from an initially empty feedback arc set $\Lambda$, in each decimation step of the BPD algorithm: (1) the BP equation is iterated  on all the remaining arcs of the network $G$ for a number $r$ of times (e.g., $r = 10$); (2) then the probability $\rho_{i j}$ for each remaining arc $(i, j)$ to be a feedback arc is estimated based on the RS mean-field formula; (3) then a tiny fraction $\epsilon$ of the remaining arcs (e.g., $\epsilon=0.01$ or $\epsilon=0.005$) which have the largest estimated feedback probabilities are deleted from the network and are added to the set $\Lambda$; (4) then the network is further simplified by recursively deleting all the nodes which have no outgoing arc or have no incoming arc.

  When there is no directed cycle in the remaining network, the BPD process will terminate. Then we check every arc $(i, j)$ in the set $\Lambda$ in a random order and delete it from $\Lambda$ if and only if the reduced set $\Lambda$ is still a feedback arc set. The final set $\Lambda$ is then reported.
  
  For random directed networks, our empirical results suggest that the BPD algorithm based on the relaxed model-E is much faster and also achieves slightly better solutions when compared with the BPD algorithm based on the strongly restricted model-R.

  \section{Minimizing the number of feedback arcs by simulated annealing}
  \label{sec:appSA}
  
  To perform simulated annealing on an input directed network $G$, we need to initialize the node permutation $\mathcal{P}=(v_1, v_2, \ldots, v_N)^T$ in a proper way. For this purpose, we first find all the strongly connected components (SCCs) of this network; and then we construct an initial node permutation $\mathcal{P}$ in the following iterative way: (1) randomly choose a SCC which receives no incoming arcs from other SCCs (there must be at least one such SCC); (2) put the nodes of this SCC to the not-yet occupied top positions of $\mathcal{P}$ in a random order; (3) then delete this SCC and all its outgoing arcs to other SCCs; (4) repeat the preceding three steps on another remaining SCC as long as the directed graph is not yet empty.
  
\begin{figure}
    \begin{center}
      \includegraphics[width=0.45\textwidth]{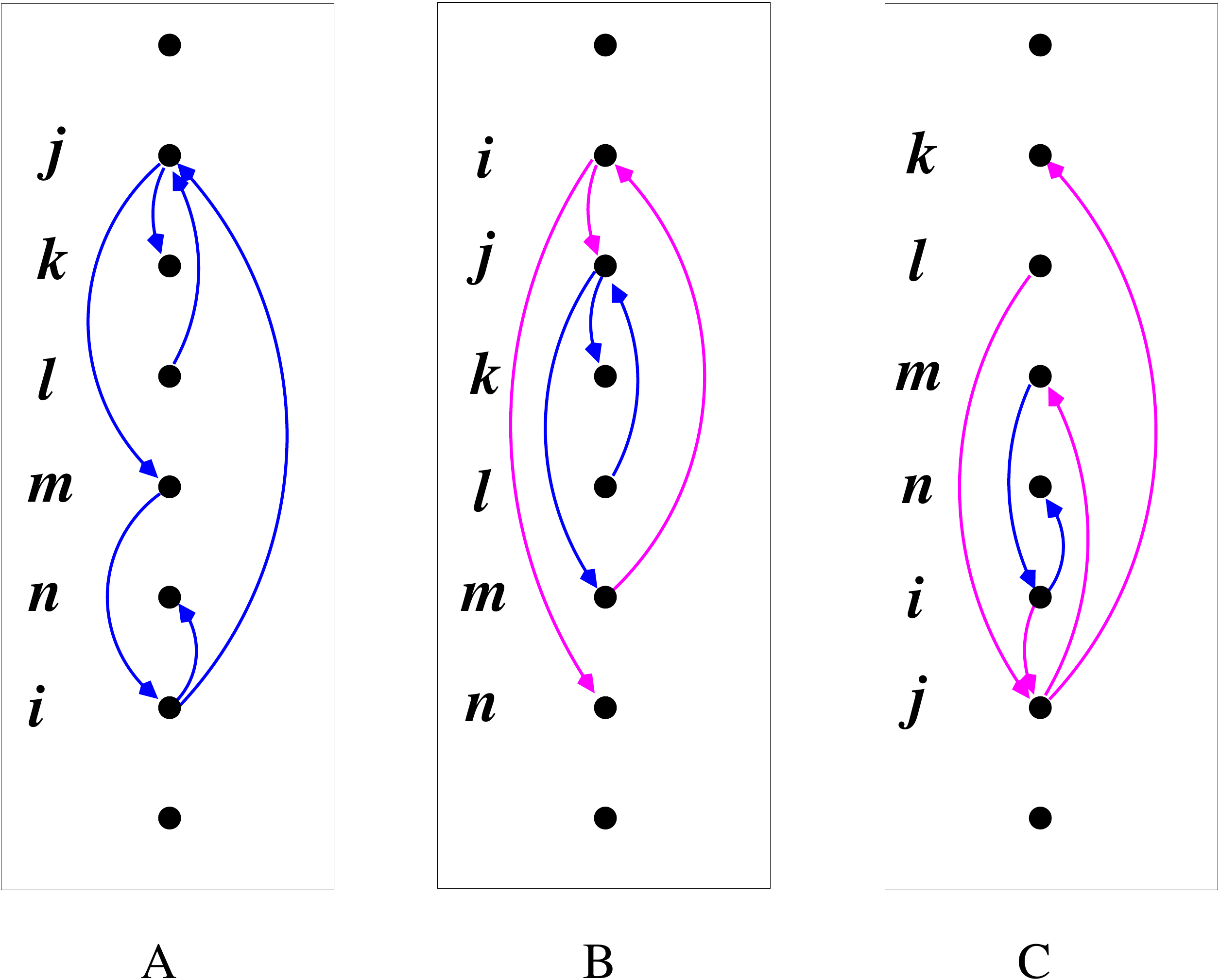}
    \end{center}
    \caption{
      \label{fig:SA}
      Changing the vertical positions of the nodes by simulated annealing to reduce the number of feedback (upward) arcs. In this example, node $i$ has two outgoing arcs $(i,j)$ and $(i,n)$ and one incoming arc $(m, i)$, while node $j$ has two outgoing arcs $(j, k)$ and $(j,m)$ and two incoming arcs $(i,j)$ and $(l, j)$. (A) The node order before updating (only the nodes $i$, $j$ and their incoming and outgoing neighbors are shown here for clarity). (B) Node $i$ is moved to the position immediately above node $j$;  the change in the number of feedback arcs is $s_{i, j}^{i \uparrow} = -1$, since now $(i, j)$ and $(i, n)$ both become feedforward (downward) arcs and  $(m,i)$ changes to be a feedback arc. (C) Node $j$ is moved to the position immediately below node $i$; the change in the number of feedback arcs is $s_{i, j}^{j\downarrow}=0$, since now $(i, j)$ and $(l,j)$ change to be feedforward arcs and $(j, k)$ and $(j, m)$ change to be feedback arcs.
    }
  \end{figure}

  We illustrate in Fig.~\ref{fig:SA} the two basic updating rules of the SA algorithm, which were inspired by the earlier work of \cite{Galinier-Lemamou-Bouzidi-2013}. This figure complements the descriptions in the main text.  In our actual implementation of the SA algorithm, to select a feedback arc most efficiently for the action shown in Fig.~\ref{fig:SA}(B), we store all the feeback arcs in different lists:  the feedback arcs $(i, j)$ whose associated energy changes $s_{i, j}^{i \uparrow} \leq 0$ are all stored in the list  $U_0$; the feedback arcs $(i, j)$ whose associated energy changes $s_{i, j}^{i \uparrow} = n$ ($\geq 1$) are all stored in the list $U_n$. To perform the updating shown in Fig.~\ref{fig:SA}(B), we first choose an integer value $n \geq 0$ according to the probability
  \begin{equation}
    P_{up}(n) = \frac{| U_n | e^{-\beta n}}{\sum\limits_{n^\prime \geq 0}
      | U_{n^\prime}| e^{-\beta n^\prime} } \; ,
  \end{equation}
  where $|U_n|$ denotes the length of the list $U_n$; then we choose an arc $(i,j)$ in the list $U_n$ uniformly at random and move node $i$ to be immediately above node $j$ in the permutation $\mathcal{P}$. 
  
  Similarly, to speed up the downward updating shown in Fig.~\ref{fig:SA}(C), we also store all the feedback arcs in another set of lists: the feedback arcs $(i, j)$ whose associated energy changes $s_{i, j}^{j \downarrow} \leq 0$ are all stored in the list  $D_0$; the feedback arcs $(i, j)$ whose associated energy changes $s_{i, j}^{j \downarrow} = n$ ($\geq 1$) are all stored in the list $D_n$. An downward updating is achieved in two steps: first, an integer value $n \geq 0$ is chosen according to the probability
  \begin{equation}
    P_{down}(n) = \frac{| D_n | e^{-\beta n}}{\sum\limits_{n^\prime \geq 0}
      | D_{n^\prime}| e^{-\beta n^\prime} } \; ,
  \end{equation}
  where $|D_n|$ denotes the length of the list $D_n$; second,  an arc $(i, j)$ in the list $D_n$ is randomly chosen and node $j$ is moved to be immediately below node $i$ in the permutation $\mathcal{P}$.

  Notice that, after each elementary updating of the node order, the  lists $U_0$, $U_1$, $\ldots$, $D_0$, $D_1$, $\ldots$ should be updated if necessary. These can be done very efficiently, since only the attached incoming and outgoing arcs of the nodes $i$ and $j$ of the chosen arc $(i, j)$ need to be considered.

  At each inverse temperature $\beta$, a total number of $2 c_0 N$ elementary updates are carried out, with one half of them being the upward movements shown in Fig.~\ref{fig:SA}(B) and the other half being the downward movements shown in Fig.~\ref{fig:SA}(C). The parameter $c_0$ is chosen in the range $c_0 \in [5, 10000]$ (the default value is $c_0=5$). Larger values of $c_0$ usually lead to slightly improved solutions, but the simulation times are longer.

  The inverse temperature $\beta$ then increases by a factor of $1/c_1$ ($c_1$ is set to be $0.99$ in the present work) after $2 c_0 N$ elementary updates have been carried out. If at $c_2$ consecutive values of $\beta$ (with $c_2=50$) the SA search process fails to find a node permutation whose corresponding number of feedback (upward) arcs is smaller than that of any previously visited node permutation, the search is then terminated and the best solution reached during the whole SA process is reported.

  \section{Random network instances}
  \label{sec:appNetwork}
  
  We generate directed random networks following conventional methods in the literature.
  
  To generate a directed random Erd\"os-R\'enyi (ER) network $G$, we start from an empty network containing $N$ nodes and then add arcs to the network one by one until the total number of added arcs reaches the specified value $M$. The two end nodes $i$ and $j$ of each candidate arc $(i, j)$ are chosen uniformly at random from the $N$ nodes, and this candidate arc is accepted if it has yet been added to the network and $i\neq j$.
  
  On the other hand, to generate a directed regular random (RR) network of arc density $\alpha$, we first assign to each node $d = 2 \alpha$ half-edges ($d$ must be an integer), and then  repeat the process of glueing two randomly chosen half-edges into a complete edge between two nodes and then assigning a random direction to this edge (self-connections and multiple arcs between the same pair of nodes are not allowed).

  To generate a directed random scale-free network according to the configurational model (i.e., a SFC network), we first assign to each node $i$ $d_i^{in}$ incoming half-arcs and $d_i^{out}$ outgoing half-arcs, with $d_i^{in}$ and $d_i^{out}$ being independent random integers generated according to the in-degree power-law distribution $P_{in}(d)$ and out-degree power-law distribution $P_{out}(d)$, respectively. The expressions for these two degree distributions are
  \begin{subequations}
    \begin{align}
      P_{in}(d)  & = \frac{d^{-\gamma_{in}}}{\sum\limits_{d^\prime = d_{min}}^{d_{max}}
        (d^\prime)^{-\gamma_{in}}} \; ,
      \quad\quad\quad ( d_{min} \leq d \leq d_{max}) \; \\
      P_{out}(d)  & = \frac{d^{-\gamma_{out}}}{\sum\limits_{d^\prime = d_{min}}^{d_{max}}
    (d^\prime)^{-\gamma_{out}}} \; , 
      \quad\quad\quad ( d_{min} \leq d \leq d_{max}) \; 
    \end{align}
  \end{subequations}
  where $\gamma_{in}$ is the in-degree decay exponent and $\gamma_{out}$ is the out-degree decay exponent; the parameter $d_{min}$ is the minimum value of degree, and $d_{max}$ is the maximum value of degree. We set $d_{min}=2$ and $d_{max}=\sqrt{N}$ in this study. After the in- and out-degree for each node are assigned, if the sum $M_{in} \equiv \sum_{i=1}^{N} d_i^{in}$ is larger (respectively, smaller) than the sum $M_{out} \equiv \sum_{i=1}^{N} d_i^{out}$, we then change $|(M_{in}-M_{out})|/2$ randomly chosen incoming (respectively, outgoing) half-arcs into outgoing (respectively, incoming) half-arcs to make $M_{in}=M_{out}$. Finally we repeat the process of glueing one randomly chosen outgoing half-arc with one randomly chosen incoming half-arc to form a complete arc (self-connections and multiple arcs between the same pair of nodes are not allowed).
  
  To generate a directed random scale-free network according to the static model (i.e., a SFS network \cite{Goh-Kahng-Kim-2001}), we proceed as follows: (1) randomly permute the $N$ nodes and then assign the $r$-th node (say node $i$) in this permutation the weight $w_i^{in}=r^{-1/(\gamma_{in}-1)}$; (2) repeatedly add an incoming half-arc to the network and attach it to a node $i$ with probability proportional to its assigned weight $w_i^{in}$; (3) randomly permute the $N$ nodes again and then assign the $r$-th node (say node $i^\prime$) in this new permutation the weight $w_{i^\prime}^{out} = r^{-1/(\gamma_{out}-1)}$;(4) repeatedly add an outgoing half-arc to the network and attach it to a node $j$ with probability proportional to its assigned weight $w_j^{out}$; (5) repeatedly glueing one randomly chosen outgoing half-arc with a randomly chosen incoming half-arc to form a full arc (self-connections and multiple arcs between the same pair of nodes are not allowed). After this directed network is generated, the probability that a randomly chosen node to have $d$ incoming arcs is $P_{in}(d) \propto d^{-\gamma_{in}}$ for large values of $d$, and the probability that a randomly chosen node to have $d$ outgoing arcs is $P_{out}(d) \propto d^{-\gamma_{out}}$ for large values of $d$ \cite{Goh-Kahng-Kim-2001}.

  \section{Visualizing a strongly connected network component}
  \label{sec:appDraw}
  
  Let us denote by $G_{SCC}$ a strongly connected component of a network $G$ and assume that this component has $n$ nodes. To visualize this strongly connected component, we need to specify for each node a two-dimensional coordinate $(x, y)$. Our recipe for this task proceeds as follows:
  \begin{enumerate}
  \item[1.] Run the SA algorithm or the BPD algorithm on $G_{SCC}$ to obtain a near-minimum feedback arc set $\Lambda$.
  \item[2.] Then construct a node hierarchy  based on set $\Lambda$ according to the simple procedure of Appendix~\ref{sec:appMap}; set the $y$-coordinate $y_i$ of each node $i$ to be identical to its hierarchical level $h_i$.
  \item[3.] Then randomly permute the $n$ nodes and assign the $r$-th node (say $i$) in this permutation the $x$-coordinate $x_i=r$.
  \item[4.] Then try to exchange the $x$-coordinates of the $n$ nodes to minimize the total ``connection cost'' $C(x_1, x_2, \ldots, x_n)$ defined by
    \begin{equation}
      \label{eq:Cost}
      C(x_1, x_2, \ldots, x_n) = \sum\limits_{(i, j)\in G_{SCC}}
      \bigl| x_i - x_j| \; .
    \end{equation}
    We can simply use simulated annealing to minimize this connection cost: In each elementary updating step, a proposal is made to exchange the $x$-coordinates of two nodes; if the change in the connection cost $\Delta C$ is zero or negative, this proposal is accepted, otherwise it is accepted with probability $e^{- \Delta C/T}$, with the parameter $T$ (the ``temperature'') being a slowly decreasing quantity.
  \item[5.] After a near-minimum $x$-coordinate solution has been obtained for the connection cost $C(x_1, \ldots, x_n)$, then draw the nodes and arcs of the strongly connected component $G_{SCC}$ according to the determined $(x, y)$ coordinates of all the $n$ nodes.
  \end{enumerate}

  The motivation for us to minimize the connection cost (\ref{eq:Cost}) is simple: we want to make the densely connected nodes to form clusters along the horizontal ($x$) direction, so that possible community structures with the strongly connected network component can be clearly visualized, besides the hierarchical structure along the vertical ($y$) direction. 

  Figure~\ref{fig:Florida} of the main text was drawn by the above-mentioned method. Notice that our method can also be applied on the whole directed network $G$: we can first work on the individual strongly connected components and then combine all these components to form a two-dimensional plot for the whole directed network.

\end{appendix}


\begin{thebibliography}{10}
\expandafter\ifx\csname url\endcsname\relax
  \def\url#1{\texttt{#1}}\fi
\expandafter\ifx\csname urlprefix\endcsname\relax\def\urlprefix{URL }\fi
\providecommand{\bibinfo}[2]{#2}
\providecommand{\eprint}[2][]{\url{#2}}

\bibitem{White-etal-1986}
\bibinfo{author}{White, J.~G.}, \bibinfo{author}{Southgate, E.},
  \bibinfo{author}{Thomson, J.~N.} \& \bibinfo{author}{Brenner, S.}
\newblock \bibinfo{title}{The structure of the nervous system of the nematode
  \textit{Caenorhabditis Elegans}}.
\newblock \emph{\bibinfo{journal}{Phil. Trans. R. Soc. Lond. B}}
  \textbf{\bibinfo{volume}{314}}, \bibinfo{pages}{1--340}
  (\bibinfo{year}{1986}).

\bibitem{Li-etal-2004}
\bibinfo{author}{Li, F.}, \bibinfo{author}{Long, T.}, \bibinfo{author}{Lu, Y.},
  \bibinfo{author}{Ouyang, Q.} \& \bibinfo{author}{Tang, C.}
\newblock \bibinfo{title}{The yeast cell-cycle network is robustly designed}.
\newblock \emph{\bibinfo{journal}{Proc. Natl. Acad. Sci. USA}}
  \textbf{\bibinfo{volume}{101}}, \bibinfo{pages}{4781--4786}
  (\bibinfo{year}{2004}).

\bibitem{Oda-etal-2005}
\bibinfo{author}{Oda, K.}, \bibinfo{author}{Matsuoka, Y.},
  \bibinfo{author}{Funahashi, A.} \& \bibinfo{author}{Kitano, H.}
\newblock \bibinfo{title}{A comprehensive pathway map of epidermal growth
  factor receptor signaling}.
\newblock \emph{\bibinfo{journal}{Mol. Syst. Biol.}}
  \textbf{\bibinfo{volume}{1}}, \bibinfo{pages}{2005.0010}
  (\bibinfo{year}{2005}).

\bibitem{Alon-2007}
\bibinfo{author}{Alon, U.}
\newblock \bibinfo{title}{Network motifs: theory and experimental approaches}.
\newblock \emph{\bibinfo{journal}{Nature Rev. Genetics}}
  \textbf{\bibinfo{volume}{8}}, \bibinfo{pages}{450--461}
  (\bibinfo{year}{2007}).

\bibitem{Milo-etal-2002}
\bibinfo{author}{Milo, R.} \emph{et~al.}
\newblock \bibinfo{title}{Network motifs: Simple building blocks of complex
  networks}.
\newblock \emph{\bibinfo{journal}{Science}} \textbf{\bibinfo{volume}{298}},
  \bibinfo{pages}{824--827} (\bibinfo{year}{2002}).

\bibitem{Leicht-Newman-2008}
\bibinfo{author}{Leicht, E.~A.} \& \bibinfo{author}{Newman, M. E.~J.}
\newblock \bibinfo{title}{Community structure in directed networks}.
\newblock \emph{\bibinfo{journal}{Phys. Rev. Lett.}}
  \textbf{\bibinfo{volume}{100}}, \bibinfo{pages}{118703}
  (\bibinfo{year}{2008}).

\bibitem{Fortunato-2010}
\bibinfo{author}{Fortunato, S.}
\newblock \bibinfo{title}{Community detection in graphs}.
\newblock \emph{\bibinfo{journal}{Phys. Rep.}} \textbf{\bibinfo{volume}{486}},
  \bibinfo{pages}{75--174} (\bibinfo{year}{2010}).

\bibitem{Tarjan-1972}
\bibinfo{author}{Tarjan, R.~E.}
\newblock \bibinfo{title}{Depth-first search and linear graph algorithms}.
\newblock \emph{\bibinfo{journal}{SIAM J. Comput.}}
  \textbf{\bibinfo{volume}{1}}, \bibinfo{pages}{146--160}
  (\bibinfo{year}{1972}).

\bibitem{Dorogovtsev-etal-2001}
\bibinfo{author}{Dorogovtsev, S.~N.}, \bibinfo{author}{Mendes, J. F.~F.} \&
  \bibinfo{author}{Samukhin, A.~N.}
\newblock \bibinfo{title}{Giant strongly connected component of directed
  networks}.
\newblock \emph{\bibinfo{journal}{Phys. Rev. E}} \textbf{\bibinfo{volume}{64}},
  \bibinfo{pages}{025101(R)} (\bibinfo{year}{2001}).

\bibitem{Jeong-etal-2000}
\bibinfo{author}{Jeong, H.}, \bibinfo{author}{Tombor, B.},
  \bibinfo{author}{Albert, R.}, \bibinfo{author}{Oltvai, Z.~N.} \&
  \bibinfo{author}{Barab\'{a}si, A.-L.}
\newblock \bibinfo{title}{The large-scale organization of metabolic networks}.
\newblock \emph{\bibinfo{journal}{Nature}} \textbf{\bibinfo{volume}{407}},
  \bibinfo{pages}{651--654} (\bibinfo{year}{2000}).

\bibitem{Ravasz-etal-2002}
\bibinfo{author}{Ravasz, E.}, \bibinfo{author}{Somera, A.~L.},
  \bibinfo{author}{Mongru, D.~A.}, \bibinfo{author}{Oltvai, Z.~N.} \&
  \bibinfo{author}{Barab{\'{a}}si, A.}
\newblock \bibinfo{title}{Hierarchical organization of modularity in metabolic
  networks}.
\newblock \emph{\bibinfo{journal}{Science}} \textbf{\bibinfo{volume}{297}},
  \bibinfo{pages}{1551--} (\bibinfo{year}{2002}).

\bibitem{Lan-Mezic-2011}
\bibinfo{author}{Lan, Y.} \& \bibinfo{author}{Mezi\'{c}, I.}
\newblock \bibinfo{title}{On the architecture of cell regulation networks}.
\newblock \emph{\bibinfo{journal}{BMC Syst. Biol.}}
  \textbf{\bibinfo{volume}{5}}, \bibinfo{pages}{37} (\bibinfo{year}{2011}).

\bibitem{CorominasMurtra-etal-2013}
\bibinfo{author}{{Corominas-Murtra}, B.}, \bibinfo{author}{Go{\~n}i, J.},
  \bibinfo{author}{Sol\'e, R.~V.} \& \bibinfo{author}{{Rodr\'iguez-Caso}, C.}
\newblock \bibinfo{title}{On the origins of hierarchy in complex networks}.
\newblock \emph{\bibinfo{journal}{Proc. Natl. Acad. Sci. USA}}
  \textbf{\bibinfo{volume}{110}}, \bibinfo{pages}{13316--13321}
  (\bibinfo{year}{2013}).

\bibitem{DominguezGarcia-Pgolotti-Munoz-2014}
\bibinfo{author}{{Dom\'inguez-Garc\'ia}, V.}, \bibinfo{author}{Pigolotti, S.}
  \& \bibinfo{author}{Mu{\~n}oz, M.~A.}
\newblock \bibinfo{title}{Inherent directionality explains the lack of feedback
  loops in empirical networks}.
\newblock \emph{\bibinfo{journal}{Sci. Rep.}} \textbf{\bibinfo{volume}{4}},
  \bibinfo{pages}{7497} (\bibinfo{year}{2014}).

\bibitem{Fiedler-etal-2013}
\bibinfo{author}{Fiedler, B.}, \bibinfo{author}{Mochizuki, A.},
  \bibinfo{author}{Kurosawa, G.} \& \bibinfo{author}{Saito, D.}
\newblock \bibinfo{title}{Dynamics and control at feedback vertex sets. i:
  Informative and determining nodes in regulatory networks}.
\newblock \emph{\bibinfo{journal}{J. Dynam. Differ. Equat.}}
  \textbf{\bibinfo{volume}{25}}, \bibinfo{pages}{563--604}
  (\bibinfo{year}{2013}).

\bibitem{Xu-Lan-2015}
\bibinfo{author}{Xu, J.} \& \bibinfo{author}{Lan, Y.}
\newblock \bibinfo{title}{Hierarchical feedback modules and reaction hubs in
  cell signaling networks}.
\newblock \emph{\bibinfo{journal}{PLoS ONE}} \textbf{\bibinfo{volume}{10(5)}},
  \bibinfo{pages}{e0125886} (\bibinfo{year}{2015}).

\bibitem{Ulanowicz-etal-1998}
\bibinfo{author}{Ulanowicz, R.~E.}, \bibinfo{author}{Bondavalli, C.} \&
  \bibinfo{author}{Egnotovich, M.~S.}
\newblock \bibinfo{title}{Network analysis of trophic dynamics in south florida
  ecosystem, fy 97: The florida bay ecosystem}.
\newblock \bibinfo{type}{Tech. Rep.}, \bibinfo{institution}{Chesapeake
  Biological Laboratory, Solomons} (\bibinfo{year}{1998}).

\bibitem{Liu-Barabasi-2016}
\bibinfo{author}{Liu, Y.-Y.} \& \bibinfo{author}{Barab\'asi, A.-L.}
\newblock \bibinfo{title}{Control principles of complex systems}.
\newblock \emph{\bibinfo{journal}{Rev. Mod. Phys.}}
  \textbf{\bibinfo{volume}{88}}, \bibinfo{pages}{035006}
  (\bibinfo{year}{2016}).

\bibitem{Eades-Sugiyama-1990}
\bibinfo{author}{Eades, P.} \& \bibinfo{author}{Sugiyama, K.}
\newblock \bibinfo{title}{How to draw a directed graph}.
\newblock \emph{\bibinfo{journal}{J. Information Processing}}
  \textbf{\bibinfo{volume}{13}}, \bibinfo{pages}{424--437}
  (\bibinfo{year}{1990}).

\bibitem{Garey-Johnson-1979}
\bibinfo{author}{Garey, M.} \& \bibinfo{author}{Johnson, D.~S.}
\newblock \emph{\bibinfo{title}{Computers and Intractability: A Guide to the
  Theory of NP-Completeness}} (\bibinfo{publisher}{Freeman},
  \bibinfo{address}{San Francisco}, \bibinfo{year}{1979}).

\bibitem{Mezard-Montanari-2009}
\bibinfo{author}{M{\'{e}}zard, M.} \& \bibinfo{author}{Montanari, A.}
\newblock \emph{\bibinfo{title}{Information, Physics, and Computation}}
  (\bibinfo{publisher}{Oxford Univ. Press}, \bibinfo{address}{New York},
  \bibinfo{year}{2009}).

\bibitem{Mezard-Parisi-2001}
\bibinfo{author}{M{\'{e}}zard, M.} \& \bibinfo{author}{Parisi, G.}
\newblock \bibinfo{title}{The bethe lattice spin glass revisited}.
\newblock \emph{\bibinfo{journal}{Eur. Phys. J. B}}
  \textbf{\bibinfo{volume}{20}}, \bibinfo{pages}{217--233}
  (\bibinfo{year}{2001}).

\bibitem{Bayati-etal-2008}
\bibinfo{author}{Bayati, M.} \emph{et~al.}
\newblock \bibinfo{title}{Statistical mechanics of steiner trees}.
\newblock \emph{\bibinfo{journal}{Phys. Rev. Lett.}}
  \textbf{\bibinfo{volume}{101}}, \bibinfo{pages}{037208}
  (\bibinfo{year}{2008}).

\bibitem{Altarelli-Braunstein-DallAsta-Zecchina-2013}
\bibinfo{author}{Altarelli, F.}, \bibinfo{author}{Braunstein, A.},
  \bibinfo{author}{{Dall'Asta}, L.} \& \bibinfo{author}{Zecchina, R.}
\newblock \bibinfo{title}{Optimizing spread dynamics on graphs by message
  passing}.
\newblock \emph{\bibinfo{journal}{J. Stat. Mech.: Theor. Exp.}}
  \bibinfo{pages}{P09011} (\bibinfo{year}{2013}).

\bibitem{Guggiola-Semerjian-2015}
\bibinfo{author}{Guggiola, A.} \& \bibinfo{author}{Semerjian, G.}
\newblock \bibinfo{title}{Minimal contagious sets in random regular graphs}.
\newblock \emph{\bibinfo{journal}{J. Stat. Phys.}}
  \textbf{\bibinfo{volume}{158}}, \bibinfo{pages}{300--358}
  (\bibinfo{year}{2015}).

\bibitem{Zhou-2016b}
\bibinfo{author}{Zhou, H.-J.}
\newblock \bibinfo{title}{A spin glass approach to the directed feedback vertex
  set problem}.
\newblock \emph{\bibinfo{journal}{J. Stat. Mech.: Theor. Exp.}}
  \bibinfo{pages}{073303} (\bibinfo{year}{2016}).

\bibitem{Kirkpatrick-etal-1983}
\bibinfo{author}{Kirkpatrick, S.}, \bibinfo{author}{{Gelatt Jr.}, C.~D.} \&
  \bibinfo{author}{Vecchi, M.~P.}
\newblock \bibinfo{title}{Optimization by simulated annealing}.
\newblock \emph{\bibinfo{journal}{Science}} \textbf{\bibinfo{volume}{220}},
  \bibinfo{pages}{671--680} (\bibinfo{year}{1983}).

\bibitem{Galinier-Lemamou-Bouzidi-2013}
\bibinfo{author}{Galinier, P.}, \bibinfo{author}{Lemamou, E.} \&
  \bibinfo{author}{Bouzidi, M.~W.}
\newblock \bibinfo{title}{Applying local search to the feedback vertex set
  problem}.
\newblock \emph{\bibinfo{journal}{J. Heuristics}}
  \textbf{\bibinfo{volume}{19}}, \bibinfo{pages}{797--818}
  (\bibinfo{year}{2013}).

\bibitem{Qin-Zhou-2014}
\bibinfo{author}{Qin, S.-M.} \& \bibinfo{author}{Zhou, H.-J.}
\newblock \bibinfo{title}{Solving the undirected feedback vertex set problem by
  local search}.
\newblock \emph{\bibinfo{journal}{Eur. Phys. J. B}}
  \textbf{\bibinfo{volume}{87}}, \bibinfo{pages}{273} (\bibinfo{year}{2014}).

\bibitem{Dorogovtsev-Mendes-2002}
\bibinfo{author}{Dorogovtsev, S.~N.} \& \bibinfo{author}{Mendes, J. F.~F.}
\newblock \bibinfo{title}{Evolution of networks}.
\newblock \emph{\bibinfo{journal}{Adv. Phys.}} \textbf{\bibinfo{volume}{51}},
  \bibinfo{pages}{1079--1187} (\bibinfo{year}{2002}).

\bibitem{Goh-Kahng-Kim-2001}
\bibinfo{author}{Goh, K.-I.}, \bibinfo{author}{Kahng, B.} \&
  \bibinfo{author}{Kim, D.}
\newblock \bibinfo{title}{Universal behavior of load distribution in scale-free
  networks}.
\newblock \emph{\bibinfo{journal}{Phys. Rev. Lett.}}
  \textbf{\bibinfo{volume}{87}}, \bibinfo{pages}{278701}
  (\bibinfo{year}{2001}).

\bibitem{Pardalos-Qian-Resende-1999}
\bibinfo{author}{Pardalos, P.~M.}, \bibinfo{author}{Qian, T.-B.} \&
  \bibinfo{author}{Resende, M. G.~C.}
\newblock \bibinfo{title}{A greedy randomized adaptive search procedure for the
  feedback vertex set problem}.
\newblock \emph{\bibinfo{journal}{J. Combin. Optim.}}
  \textbf{\bibinfo{volume}{2}}, \bibinfo{pages}{399--412}
  (\bibinfo{year}{1999}).

\bibitem{Leskovec-Huttenlocher-Kleinberg-2010-b}
\bibinfo{author}{Leskovec, J.}, \bibinfo{author}{Huttenlocher, D.} \&
  \bibinfo{author}{Kleinberg, J.}
\newblock \bibinfo{title}{Predicting positive and negative links in online
  social networks}.
\newblock In \emph{\bibinfo{booktitle}{Proceedings of the 19th International
  Conference on World Wide Web}}, \bibinfo{pages}{641--650}
  (\bibinfo{publisher}{ACM}, \bibinfo{address}{New York},
  \bibinfo{year}{2010}).

\bibitem{Ripeanu-etal-2002}
\bibinfo{author}{Ripeanu, M.}, \bibinfo{author}{Foster, I.} \&
  \bibinfo{author}{Iamnitchi, A.}
\newblock \bibinfo{title}{Mapping the gnutella network: Properties of
  large-scale peer-to-peer systems and implications for system design}.
\newblock \emph{\bibinfo{journal}{IEEE Internet Comput.}}
  \textbf{\bibinfo{volume}{6}}, \bibinfo{pages}{50--57} (\bibinfo{year}{2002}).

\bibitem{Mugisha-Zhou-2016}
\bibinfo{author}{Mugisha, S.} \& \bibinfo{author}{Zhou, H.-J.}
\newblock \bibinfo{title}{Identifying optimal targets of network attack by
  belief propagation}.
\newblock \emph{\bibinfo{journal}{Phys. Rev. E}} \textbf{\bibinfo{volume}{94}},
  \bibinfo{pages}{012305} (\bibinfo{year}{2016}).

\bibitem{Braunstein-etal-2016}
\bibinfo{author}{Braunstein, A.}, \bibinfo{author}{Dall'Asta, L.},
  \bibinfo{author}{Semerjian, G.} \& \bibinfo{author}{Zdeborov\'a, L.}
\newblock \bibinfo{title}{Network dismantling}.
\newblock \emph{\bibinfo{journal}{Proc. Natl. Acad. Sci. USA}}
  \textbf{\bibinfo{volume}{113}}, \bibinfo{pages}{12368--12373}
  (\bibinfo{year}{2016}).

\end{thebibliography}
\end{document}